\documentclass[useAMS,useAMSmath,usenatbib,numberedappendix]{emulateapj}

\usepackage{comment}
\usepackage{xspace}

\newcommand{\sgra}{Sgr A$^*$\xspace}

\shorttitle{X-ray Flux Distribution of \sgra}
\shortauthors{Neilsen et al.}

\begin{document}

\title{The X-ray Flux Distribution of Sagittarius A* as Seen by \textit{Chandra}}

\author{J.\ Neilsen\altaffilmark{1,2}, S.\ Markoff\altaffilmark{3}, M.~A.\ Nowak\altaffilmark{2}, J.\ Dexter\altaffilmark{4}, G.\ Witzel\altaffilmark{5}, N.\ Barri\`ere\altaffilmark{6}, Y.\ Li\altaffilmark{7}, F.~K.\ Baganoff\altaffilmark{2}, N.\ Degenaar\altaffilmark{8}, P.~C.\ Fragile\altaffilmark{9}, C.\ Gammie\altaffilmark{10,11}, A.\ Goldwurm\altaffilmark{12,13}, N.\ Grosso\altaffilmark{14},  D.\ Haggard\altaffilmark{15}} 
\altaffiltext{1}{Einstein Fellow, Boston University Department of Astronomy, Boston, MA 02215, USA; neilsenj@bu.edu}
\altaffiltext{2}{MIT Kavli Institute for Astrophysics and Space Research, Cambridge, MA 02139, USA}
\altaffiltext{3}{Astronomical Institute, ``Anton Pannekoek'', University of Amsterdam, Postbus 94249, 1090 GE Amsterdam, The Netherlands}
\altaffiltext{4}{Department of Astronomy, Hearst Field Annex, University of California, Berkeley, CA 94720-3411, USA}
\altaffiltext{5}{Department of Physics and Astronomy, University of California, Los Angeles, CA 90095-1547, USA}
\altaffiltext{6}{Space Sciences Laboratory, 7 Gauss Way, University of California, Berkeley, CA 94720-7450}
\altaffiltext{7}{Department of Astronomy and Institute of Theoretical Physics and Astrophysics, Xiamen University, Xiamen, Fujian 361005, China}
\altaffiltext{8}{Hubble Fellow, Department of Astronomy, University of Michigan, 500 Church Street, Ann Arbor, MI 48109, USA}
\altaffiltext{9}{Department of Physics \& Astronomy, College of Charleston, Charleston, SC 29424, USA}
\altaffiltext{10}{Department of Astronomy, University of Illinois Urbana-Champaign, 1002 W. Green St., Urbana, IL 61801, USA}
\altaffiltext{11}{Department of Physics, University of Illinois Urbana-Champaign, 1110 W. Green St., Urbana, IL 61801, USA}
\altaffiltext{12}{AstroParticule et Cosmologie (APC), Universit\'{e} Paris 7 Denis Diderot, 75205 Paris cedex 13, France}
\altaffiltext{13}{Service d'Astrophysique/IRFU/DSM, CEA Saclay, 91191 Gif-sur-Yvette cedex, France}
\altaffiltext{14}{Observatoire Astronomique de Strasbourg, Universit\'e de Strasbourg, CNRS, UMR 7550, 11 rue de l'Universit\'e, 67000 Strasbourg, France}
\altaffiltext{15}{Center for Interdisciplinary Exploration and Research in Astrophysics, Physics and Astronomy Department, Northwestern University, Evanston, IL 60208, USA; CIERA Postdoctoral Fellow}  

\begin{abstract}
We present a statistical analysis of the X-ray flux distribution of \sgra from the \textit{Chandra X-ray Observatory}'s 3 Ms \sgra X-ray Visionary Project (XVP) in 2012. Our analysis indicates that the observed X-ray flux distribution can be decomposed into a steady quiescent component, represented by a Poisson process with rate $Q=(5.24\pm0.08)\times10^{-3}$ cts s$^{-1},$ and a variable component, represented by a power law process ($dN/dF\propto F^{-\xi},$ $\xi=1.92_{-0.02}^{+0.03}$). This slope matches our recently-reported distribution of flare luminosities. The variability may also be described by a log-normal process with a median unabsorbed 2--8 keV flux of $1.8^{+0.9}_{-0.6}\times10^{-14}$ erg s$^{-1}$ cm$^{-2}$ and a shape parameter $\sigma=2.4\pm0.2,$ but the power law provides a superior description of the data. In this decomposition of the flux distribution, all of the intrinsic X-ray variability of \sgra (spanning at least three orders of magnitude in flux) can be attributed to flaring activity, likely in the inner accretion flow. We confirm that at the faint end, the variable component contributes $\sim10\%$ of the apparent quiescent flux, as previously indicated by our statistical analysis of X-ray flares in these \textit{Chandra} observations. Our flux distribution provides a new and important observational constraint on theoretical models of \sgra, and we use simple radiation models to explore the extent to which a statistical comparison of the X-ray and infrared can provide insights into the physics of the X-ray emission mechanism.
\end{abstract}
                 

\section{INTRODUCTION}\label{sec:intro}

\defcitealias{Witzel12}{W12}
\defcitealias{Dodds-Eden11}{D11}
\setcounter{footnote}{0}

After over a decade of multiwavelength monitoring of \sgra, the supermassive black hole at the center of the Galaxy, significant uncertainties regarding the physics of its X-ray emission persist. Observationally, the X-ray emission seen by \textit{Chandra, XMM-Newton, NuSTAR,} and \textit{Swift} appears to be composed of a steady quiescent background interrupted roughly daily by flares (e.g., \citealt{Baganoff01,Baganoff03,Goldwurm03,Porquet03,Eckart04,Belanger05,Eckart06,Porquet08,Trap11,Nowak12,N13b,Degenaar13,Barriere14}). The quiescent flux is exceedingly faint, corresponding to a 2--10 keV luminosity $L_{\rm X}\approx3.5\times10^{33}$ erg s$^{-1}\lesssim10^{-11}L_{\rm Edd},$ where the Eddington luminosity $L_{\rm Edd}$ is the canonical maximum luminosity from an accreting $4.1\times10^{6}$ $M_{\sun}$ black hole. Presently (for a recent review, see \citealt{Yuan14}), the quiescent X-ray source is understood as thermal plasma emission from an accretion flow extending out to the Bondi radius (\citealt{Quataert02,Baganoff03,Yuan03,Liu04,Xu06,Wang13}). \sgra appears to owe its low luminosity to a combination of extremely radiatively inefficient accretion (\citealt{Narayan95a,Quataert00a,Yuan03,Blandford99,Yuan12}) and an outflow that removes $\sim99\%$ of the inflowing matter before it reaches the event horizon (see \citealt{Wang13} and references therein).

The origin of the X-ray flares is somewhat more elusive, a fact best illustrated by reference to the panoply of models suggested to produce them, which include particle acceleration and/or heating via magnetic reconnection or other stochastic processes in the inner accretion flow or base of an outflow (e.g., \citealt{Markoff01,Liu02,Yuan03,Liu04}), as well as the tidal vaporization of asteroids (\citealt*{Cadez08}; \citealt{Kostic09}; \citealt*{Zubovas12}). The X-rays themselves have been attributed to direct synchrotron, synchrotron self-Compton (SSC; in which infrared-emitting electrons inverse Compton scatter the infrared synchrotron photons), or other inverse Compton scenarios, in which either the IR photons are upscattered by submillimeter (sub-mm) emitting electrons or the sub-mm photons are upscattered by the IR-emitting electrons (\citealt{Markoff01,Liu02,Liu04,Yuan03,Yuan04,Eckart04,Eckart06,Marrone08,Eckart09,Dodds-Eden09,Yuan09b,Yusef-Zadeh09,Witzel12,Yusef-Zadeh12,Nowak12,Barriere14}). 

Part of the reason for the model degeneracy is the significant interstellar absorption towards the Galactic center, which makes it difficult to confirm or exclude curvature in the X-ray spectrum over the narrow bandpasses of \textit{Chandra} and \textit{XMM-Newton}. In addition, it has proven difficult to acquire truly simultaneous sub-mm, IR, and X-ray observations of a large number of flares. Presently, however, synchrotron models with cooling breaks appear to be slightly favored over inverse Compton and SSC scenarios, given the absence of significant curvature out to $\sim60$ keV in a \textit{NuSTAR} spectrum of a bright flare (\citealt{Barriere14}; see also \citealt{Dodds-Eden09,Dibi14a}). But even if the model degeneracy remains, the abundance of models reveals these flares to be astrophysically significant, by far the closest example of supermassive black hole variability. With connections to the fundamental plane (and thus AGN, blazars, and even X-ray binaries, \citealt{Merloni03,Falcke04,Markoff05,Plotkin12}), these flares may be our best opportunity to understand \sgra in the context of other accreting systems.

Thus an alternative approach may be required to provide deeper insight into the radiation physics of \sgra. For example, a multiwavelength statistical analysis that takes advantage of the wealth of available data may be a more powerful way to reveal correlations between wavelength bands and physical processes. A great deal has already been learned from statistical investigations of the Galactic center in the near infrared (NIR), where flares are at least four times more common (e.g., \citealt{Genzel03,Eckart06}). For example, \citet{Meyer09} reported a break in the NIR power spectrum at a time scale of $\sim150$ minutes, which is consistent with a linear scaling between break frequency and black hole mass in relation to the AGN studied by \citeauthor{Uttley05} (\citeyear{Uttley05}; see also \citealt{McHardy06,Meyer08,Do09,Meyer14_arxiv}). \citeauthor{Dodds-Eden11} (\citeyear{Dodds-Eden11}, hereafter \citetalias{Dodds-Eden11}) analyzed five years of VLT $Ks$-band observations of \sgra and reported that the distribution of infrared flux $F_{\rm IR}$ could be described by a log-normal distribution (Eq.\ \ref{eq:lognormal}) with a median flux of $\sim1$ mJy, breaking to a power law tail at high flux ($\gtrsim5$ mJy, $\sim F_{\rm IR}^{-2.7}$). Mirroring the X-ray band, they argued that this implies two states of infrared emission (quiescent and flaring). However, in a subsequent analysis of a larger VLT data set, \citeauthor{Witzel12} (\citeyear{Witzel12}, hereafter \citetalias{Witzel12}) found that the flux distribution was consistent with a pure power law ($\sim F_{\rm IR}^{-4.2}$). In either case, it appears that \sgra is continuously variable over long and short time scales in the NIR. 

Thanks to the 2012 \textit{Chandra} X-ray Visionary Project\footnote{http://www.sgra-star.com} on \sgra (hereafter the XVP, a 3 Ms campaign to observe the Galactic center at high spectral/spatial resolution and high cadence), we have made a number of recent advances in understanding the black hole's X-ray emission, most recently demonstrating that the quiescent emission cannot be produced by a cluster of coronally active stars (\citealt{Wang13} and references therein). The \textit{Chandra} data include the brightest known X-ray flare from \sgra (\citealt{Nowak12}), as well as dozens of others (nearly tripling the number of observed flares). Based on our analysis of the luminosity and fluence distributions of these flares (\citealt{N13b}), we estimated that (1) detectable flares constitute approximately one-third of the total X-ray emission on time scales of 3 Ms, and (2) undetected flares likely contribute $\lesssim10\%$ of the quiescent emission. It can also be demonstrated that when extrapolated to very high luminosity ($L_{\rm X}\gtrsim10^{39}$ erg s$^{-1}),$ our flare luminosity distribution is broadly consistent with the historical frequency of extreme outbursts of \sgra (e.g., \citealt{Ponti10,Clavel13}). 

In addition, in the XVP data we now have a statistical sample of X-ray fluxes that approaches what is available at NIR wavelengths. We can therefore apply similar statistical techniques as a complementary approach to observationally expensive multiwavelength campaigns; by comparing statistics between wavebands, we can search for connections and new insights into the radiation mechanisms of the black hole. In this context, we have undertaken an analysis of the X-ray flux distribution of \sgra. In this paper, we address the X-ray lightcurves from a new perspective: rather than dividing the data into ``quiescent" and ``flare" states, we suppose that the emission in each time bin is the sum of two continuous processes, one associated with the quiescent accretion flow and one associated with the flaring, active (variable) region close to the black hole. We briefly describe the data in Section \ref{sec:obs} and detail our analysis in Section \ref{sec:flux}. Flux distribution in hand, we present preliminary considerations of radiation mechanisms in light of our results and the NIR results in Section \ref{sec:radiation}, and summarize in Section \ref{sec:discuss}.

\section{Observations}
\label{sec:obs}
For the details of the \textit{Chandra} observations and our data reduction and analysis methods, we refer the reader to \citet{N13b}. Briefly, our data consist of 38 high spatial/spectral resolution \textit{Chandra} HETGS observations of \sgra from 2012. We restrict our analysis to the 2012 dataset, with its unique combination of the HETGS spectral resolution, observing cadence, and uniform calibration. From each observation, we extract 2--8 keV light curves in 300 s bins (taking detected counts from the inner $1.25\arcsec$ of the zeroth order image and the $\pm1^{\rm st}$ diffraction orders; see Figure 1 of \citealt{N13b}). The typical X-ray flare duration is $\sim3$ ks, so most detectable flares are well resolved by these data. However, as discussed in \citet{Nowak12} and \citet{N13b}, the X-ray light curve of \sgra may exhibit substructure on time scales shorter than 300 s (see also Haggard et al., in preparation), so the count rates reported here should be considered averages over 300 s, analogous to the (shorter) integration times in the NIR. We base our flux calculations (2--8 keV) on the results of \citet{Nowak12}: the quiescent unabsorbed flux is $0.45\times10^{-12}$ erg s$^{-1}$ cm$^{-2}$ and the average unabsorbed flux of the brightest flare is $21.6\times10^{-12}$ erg s$^{-1}$ cm$^{-2}$. We perform all our analysis in the Interactive Spectral Interpretation System (ISIS; \citealt{HD00,Houck02}). 

\begin{figure}
\centerline{\includegraphics[width=3.2 in]{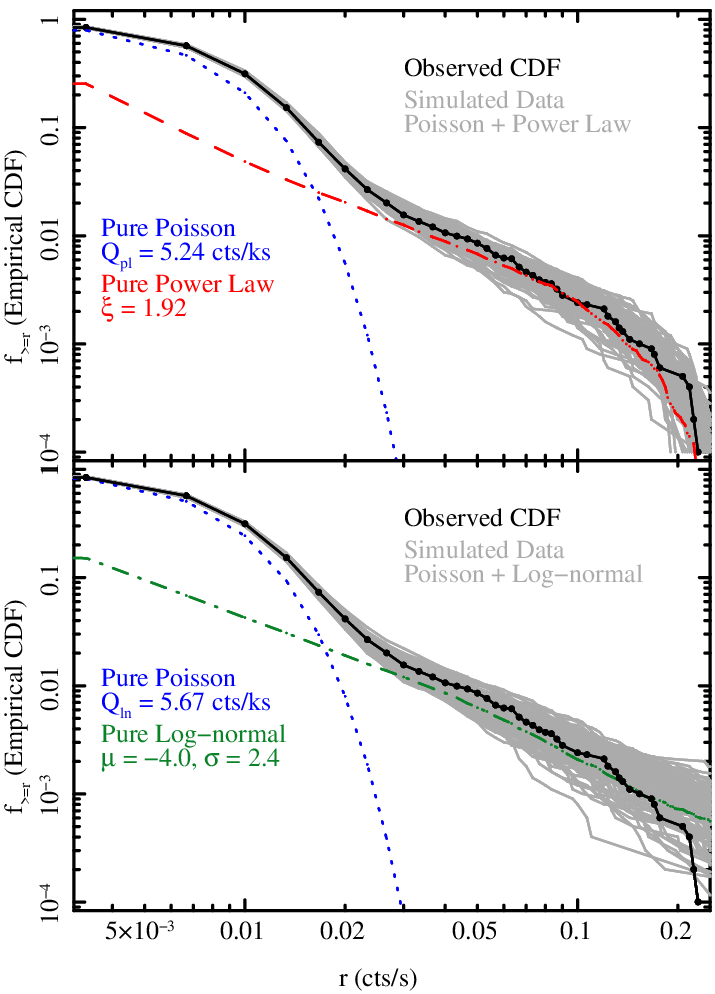}}
\caption{\textit{Chandra} X-ray count rate distribution of \sgra from 2013, defined as the fraction of time bins with a count rate greater than or equal to a given rate (black). We model this distribution as the sum of a Poisson process with rate $Q$ to represent the quiescent emission at low flux, and a variable process to represent the flare emission. In the top panel, we model the variable process as a power law with index $\xi$. The solid grey curves represent simulated data with quiescent count rates and power law indices drawn at random from the joint probability distribution of $\xi$ and $Q_{pl}$ in Figure \ref{fig:contour}. The blue dotted and red dashed curves represent sample pure Poisson and pure power law processes, respectively, with $Q_{pl}$ and $\xi$ set to their most probable values. In the bottom panel, we present a log-normal model for the variable process, with location parameter $\mu$ and shape parameter $\sigma.$ Again, the grey curves represent synthetic data, the dotted blue line represents a pure Poisson process, and the dash-dotted green line represents a pure log-normal process. See Section \ref{sec:flux} for details.\label{fig:dist}}
\end{figure}

\section{THE X-RAY FLUX DISTRIBUTION}
\label{sec:flux}

In order to determine the NIR flux distribution of \sgra, \citetalias{Witzel12} used Kolmogorov-Smirnov (K-S) tests to evaluate the goodness of fit for a range of power law indices. Here, because we are interested in the fluxes in excess of the quiescent background emission from the Galactic center, we adopt a modified version of this technique. Since we incorporate every 300 s time bin from the 3 Ms XVP campaign, our present analysis samples all flares, even those too short or too faint to be identified by our detection algorithms (\citealt{N13b}).

\subsection{Observed Flux Distribution}
Following  \citetalias{Witzel12}, we analyze a (complementary) cumulative distribution function (CDF), i.e., the fraction of count rates greater than or equal to a given rate:
\begin{equation}
\label{eq:cdf}
f_{\geq}(r)=\frac{1}{M}\sum_{i=1}^{M}{\tt if}(r_i\geq r),
\end{equation}
where $r_i$ are the count rates and $M=9964$ is the number of data points (a data point being the count rate in one 300 s time bin). X-ray count rates are preferable to flux as an observable because they are not model-dependent. The resulting empirical CDF of \sgra from 2012 is shown in black in Figure \ref{fig:dist}. For reference, the blue dotted line is the CDF for a Poisson distribution with a rate $Q\sim5.24\times10^{-3}$ cts s$^{-1}$ (see below). It is apparent that the lowest count rates ($r\lesssim0.02$ cts s$^{-1}$) are dominated by this Poisson component. Although we cannot define quiescence based on the count rate alone, this figure supports our previous estimate that $\gtrsim90\%$ of the quiescent emission can be described as a Poisson process. At higher count rates, there is also a clear excess above the quiescent Poisson noise. The sharp break at the bright end of the CDF is likely a feature of the empirical CDF itself, which must converge to $1/M$ at the highest count rate bin.

To study the intrinsic flux distribution, we must be able to predict count rates given an intrinsic flux. Since there is no conclusive evidence\footnote{Both \citet{Degenaar13} and \citet{Barriere14} present suggestive evidence for spectral variations between flares ($\sim90-95\%$ confidence), but we are not aware of any highly significant detections.} for significant variations in the flare spectrum (\citealt{Porquet08,Nowak12,N13b,Degenaar13,Barriere14}), we can use the light curve analysis from \citet{N13b} and the spectral analysis of the brightest flare from \citet{Nowak12}  to convert any unabsorbed flare flux $F$ from our models into a model flare count rate $r$ (i.e., above the quiescent background). Specifically, 
\begin{equation}
\label{eq:r}
r(F)\approx\frac{1}{\Delta t}{\tt prand}\left[\mathcal{P}\left(F~\frac{\mathcal{P}^{-1}(r_{\rm max})}{F_{\rm max}}\right)\Delta t\right]
\end{equation} where $\Delta t=300$ s is the bin time, $F_{\rm max}=21.6\times10^{-12}$ erg s$^{-1}$ cm$^{-2}$ is the mean unabsorbed flux of the brightest flare, $r_{\rm max}\sim0.13$ cts s$^{-1}$ is the mean observed count rate of the brightest flare, {\tt prand} is a Poisson random number generator to incorporate counting noise, and $\mathcal{P}$ represents the suppression of the count rate due to photon pileup (\citealt{Davis01,N13b}). The result is approximate because we have neglected the contribution of the quiescent background to pileup. For the very low count rates involved here, pileup is still fairly linear in the count rate, and we estimate that the approximation is good to within 1\%, so that any errors are at a level well below the counting noise.

\subsection{Power Law Excess}
\label{sec:pow}
To model the excess flare emission, we first consider a power law model, since a scale-free representation simplifies our investigation of the radiation mechanism (Section \ref{sec:radiation}). For an excess flux $F$ in units of 10$^{-12}$ erg s$^{-1}$ cm$^{-2}$:
\begin{equation}
P(F) = \left\{
   \begin{array}{ll}
   k F^{-\xi} & : F_1 \leq F\leq F_2\\
    0 & : \mathrm{otherwise}
   \end{array},
\right.\label{eq:powlaw}
\end{equation}
where $P(F)$ is the probability of an excess flux $F$, the lower bound $F_1=0.04$ is well below our detection limit ($F\sim0.25$), the upper bound $F_2=41$ corresponds to the maximum observed count rate of $\sim0.23$ cts s$^{-1}$, and $k=(1-\xi)/(F_1^{1-\xi}-F_2^{1-\xi})$ is a normalization constant. $F_1=0.04$ leads to the best match to our data (compared to other values between 0.01 and 0.1), and should be far enough below our detection limit and the upper limit $F_2$ that the bounded power law model is effectively scale-free in practice. We consider two ways to test this model: statistical tests using synthetic (simulated) data and maximum likelihood methods, specifically Markov Chain Monte Carlo (MCMC).

\subsubsection{Synthetic Data}
To use synthetic data, we modify the technique used by \citetalias{Witzel12}: as they did, we generate a power law random variable according to the probability distribution in Equation \ref{eq:powlaw}. Here, however, we also apply Equation \ref{eq:r} to convert fluxes to counts, then compare the sum of the power law process and a Poisson process (representing the quiescent emission, which is not seen in the NIR) to the observed data. Instead of a K-S test, we use a two-sample Anderson-Darling test (A-D; \citealt*{AndersonDarling54,Pettitt76,Scholz87}), which is more sensitive to the tails of the tested distributions (a valuable property given the proportion of values at high fluxes in our data). We refer the reader to \citet{Scholz87}, particularly their example in Section 7, for the details of calculating the test statistic in the presence of ties, as well as a table of percentile points and significance levels $p$. Outside the range of this table, we use a linear extrapolation of $\log(p/(1-p))$, similar to the implementation in the $R$ statistical package\footnote{http://www.inside-r.org/packages/cran/adk/docs/adk.test As noted there by Scholz, very large and very small $p$ values should be treated as approximate, but  ``this should not strongly affect any decisions regarding the tested hypothesis."}.

\begin{figure}
\centerline{\includegraphics[width=3.2in]{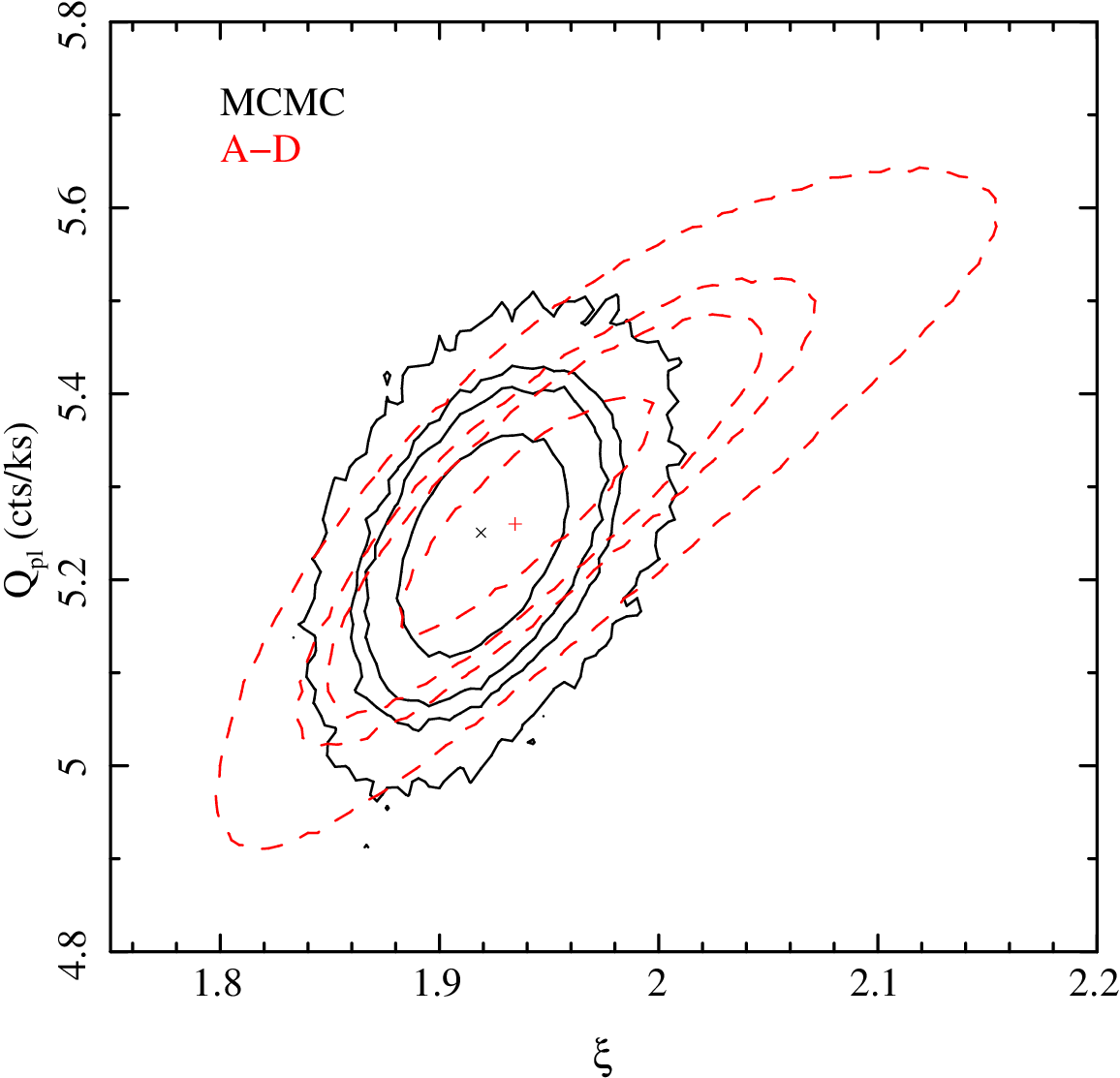}}
\caption{Probability contours in the $\xi-Q_{pl}$ plane for the power law analysis. The black contours contain 68\%, 90\%, 95\%, and 99.7\% of the ensemble of walkers in our MCMC run. The red dashed contours are lines of constant $p$-value from our Anderson-Darling tests, corresponding to confidence levels of 68\%, 90\%, 95\%, and 99.7\%. The results are weakly sensitive to the methods used.\vspace{-2mm}\label{fig:contour}}
\end{figure}

We repeat this test for a grid of 101 evenly spaced values of $\xi$ between 1.75 and 2.2 and 101 evenly spaced values of $Q_{pl}$ between 0.0048 and 0.0058 cts s$^{-1}$, inclusive (the subscript $pl$ indicates quantities found in the power law analysis). Outside these ranges, additional tests indicate that the match probabilities are negligible. For each pair ($\xi,Q_{pl}$), we generate 1000 simulated data sets including counting noise and record the average $p$-value from our A-D tests with the observed data. The contours of constant $p$-value are shown as dashed red lines in Figure \ref{fig:contour}. The highest $p$-values ($\sim0.5$) are returned in the vicinity of $Q_{pl}\sim5.26\times10^{-3}$ cts s$^{-1}$ and $\xi\sim1.93.$ Next, we confirm this model using a more detailed MCMC analysis.

\subsubsection{Maximum Likelihood}
\label{sec:pllike}
For the models considered here, in which the data are treated as the sum of two random processes with significant counting noise, the likelihood function is not trivial to write down. For details, see Appendix \ref{sec:likelihood}, but briefly the probability of observing $n$ counts is a convolution of the Poisson distribution (representing the quiescent emission) and a piled-up power law with counting noise (here representing the flare emission). The power law term can be calculated analytically, so the log-likelihood easily calculated. This makes the problem ideal for MCMC. We have implemented the parallel stretch from {\tt emcee} (\citealt{Foreman-Mackey13}) in ISIS. We use an ensemble of 300 walkers  -- 150 per free parameter -- initially distributed uniformly throughout a suitably wide region of parameter space ($1.5\leq\xi\leq2.5,$ $0.004\leq Q_{pl}~{\rm cts}^{-1}~{\rm s}\leq0.007$). The walkers are evolved for 3000 steps, and we discard the first 600 (we estimate that the average autocorrelation time is $\lesssim50$ steps). In order to keep the acceptance fraction below 0.5, we set the $a$-parameter, which determines the size of the stretch move at each step (see \citealt{Foreman-Mackey13,GoodmanWeare}), to 4. We assume uniform priors for all parameters.

\begin{figure}
\centerline{\includegraphics[width=3.2in]{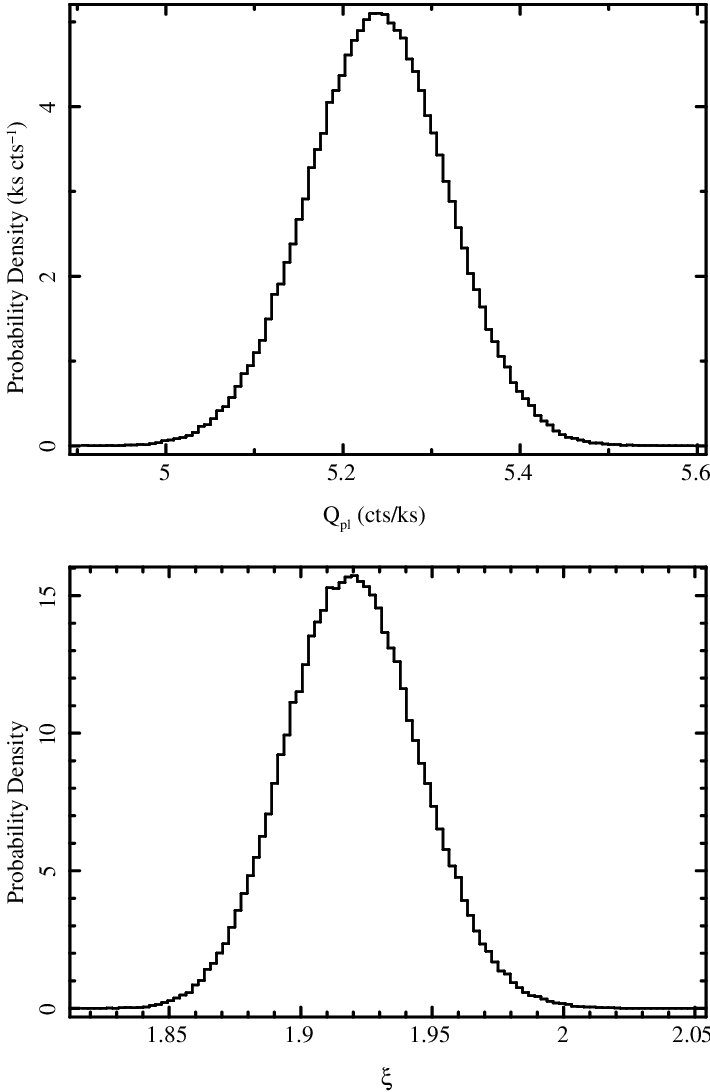}}
\caption{MCMC PDFs for the quiescent rate $Q_{pl}$ (top) and the power law index $\xi$ (bottom) marginalized over $\xi$ and $Q,$ respectively. Both distributions are well behaved. Note that as probability densities, these curves integrate to 1.\label{fig:pdf}}
\end{figure}

In Figure \ref{fig:contour}, we show the contours (black) in the $\xi-Q_{pl}$ plane that contain 68\%, 90\%, 95\%, and 99.7\% of the resulting ensemble of walkers, which agrees well with the $p$-value contours from the A-D analysis. There is clearly a small correlation between $\xi$ and $Q_{pl},$ but both parameters are well constrained and the correlation is smaller than in the A-D analysis. The probability density functions (PDFs) for $\xi$ and $Q_{pl}$ are presented in Figure \ref{fig:pdf}. We find $\xi=1.92_{-0.02}^{+0.03}$ and $Q_{pl}=(5.24\pm0.08)\times10^{-3}$ cts s$^{-1}$ (these values correspond to the 16th, 50th, and 84th percentiles of the marginalized distributions\footnote{http://dan.iel.fm/emcee/current/user/line/\#the-generative-probabilistic-model}). To demonstrate that the ensemble is able to reproduce the observed CDF, we draw 100 random ($\xi,~Q_{pl}$) pairs from the probability distribution shown in Figure \ref{fig:contour} and use them to generate simulated Poisson-plus-power law data (converting power law fluxes to counts using Equation \ref{eq:r}). The associated CDFs for the simulated datasets, shown in gray in the top panel of Figure \ref{fig:dist}, match the data very well at all count rates. Note that if we allow $F_1$ to vary as well, the result is $F_1=0.04\pm0.02$ and there is no significant change in the other parameters, although the error bars are somewhat larger. 

\subsection{Log-normal Excess}

\label{sec:lognormal}

It is worth exploring alternatives to the power law model, and the log-normal is a natural choice. Tilted accretion disk models of \sgra tend to produce NIR flux distributions that are well described by a log-normal model (\citealt{Dexter13}). A log-normal flux distribution is expected for an exponentiated damped random walk process (\citealt{Meyer14_arxiv}), and damped random walk models can successfully describe \sgra's sub-mm variability (see \citealt{Dexter14} and references therein).

Ideally, we would compare the power law model to a model of the same form as used by \citet{Dodds-Eden11}, i.e., a log-normal distribution at low flux transitioning to a power law at high flux. But since the power law already provides a satisfactory description of the bright end of the X-ray flux distribution and the faint end is dominated by the quiescent Poisson process, and the more complicated function form would have an additional three free parameters, we focus on a pure log-normal process for the time being.

We proceed as in Section \ref{sec:pow}, replacing the power law random distribution with a log-normal probability distribution: \begin{equation}
P(F) =  \frac{1}{\sqrt{2\pi}\sigma F}\exp\left(-\frac{(\ln{F}-\mu)^2}{2\sigma^2}\right).\label{eq:lognormal}\end{equation} With $F$ scaled to $10^{-12}$ erg s$^{-1}$ cm$^{-2},$ $\mu$ and $\sigma$ are dimensionless. As explained in \citet{Dodds-Eden11}, this log-normal distribution has a median flux of $\exp(\mu)\times10^{-12}$ erg s$^{-1}$ cm$^{-2}$ and a multiplicative standard deviation of $\exp(\sigma).$ 

\begin{figure}
\centerline{\includegraphics[width=3.2in]{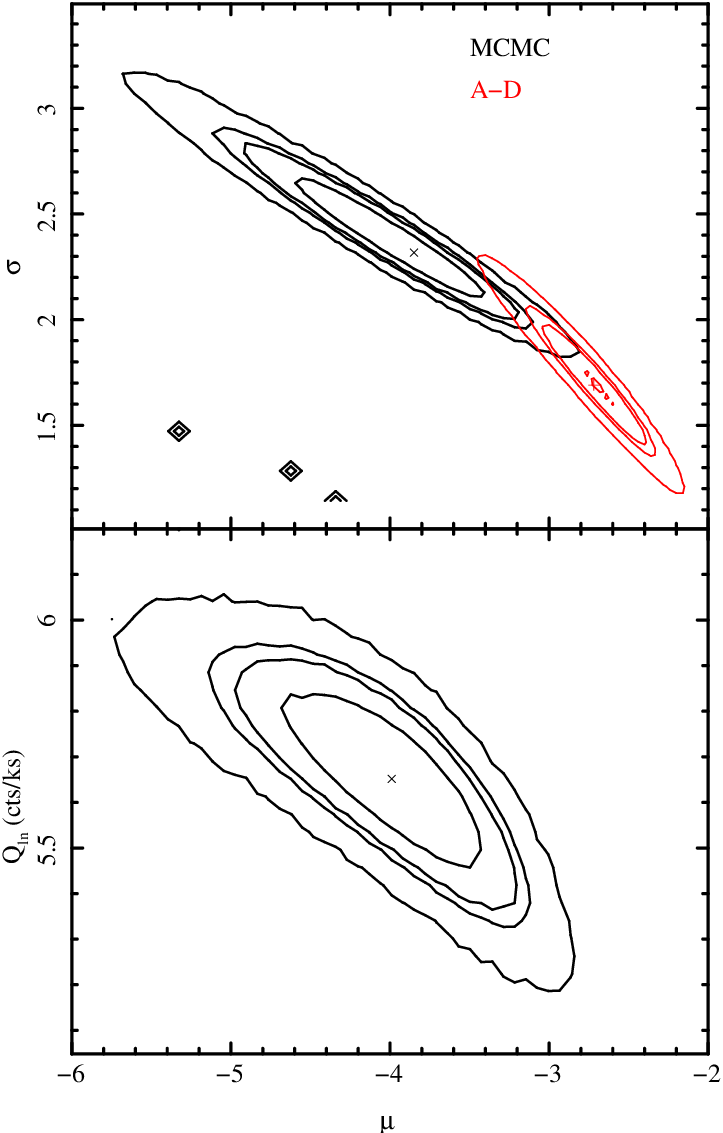}}
\caption{Probability contours in the $\mu-\sigma$ (top panel) and $\mu-Q_{ln}$ (bottom panel) planes for the log-normal analysis. Contours are as in Figure \ref{fig:contour} (black for MCMC, red for Anderson-Darling); results differ slightly between the two methods. Only MCMC contours are shown in the bottom panel, since $Q_{ln}$ is fixed in the A-D analysis. The ``diamonds" in the top panel are small regions where the probability density has a local maximum; they can also be seen in the PDFs in Figure \ref{fig:lnpdf}. \label{fig:lognom}}
\end{figure}

\subsubsection{Synthetic Data}

For the synthetic log-normal analysis, we calculate A-D test statistics over a $101\times101$ grid covering $-6\leq\mu\leq-2$ (corresponding to median fluxes for the log-normal component from $\sim2.5\times10^{-15}$ erg s$^{-1}$ cm$^{-2}$ to $\sim1.4\times10^{-13}$ erg s$^{-1}$ cm$^{-2}$), and $1\leq\sigma\leq5;$ we fix $Q$ at $Q_{ln}=0.00526$ cts s$^{-1},$ which corresponds to the peak $p$-value in the synthetic power law analysis. All the $\mu$ values tested here correspond to median fluxes for the log-normal process that lie below our detection limit. This is not a weakness of the log-normal analysis, rather it is a generic property of the high flux tail: the observed excess represents only a few percent of the data bins and only a fraction of the variable process (regardless of the model; see Figure \ref{fig:dist}). Indeed, while all the simulated fluxes in both models are positive, most are so small that the most likely discrete count rate for both the power law and the log-normal components is 0 cts s$^{-1}$. 

We present the resulting $p$-value contours in red in Figure \ref{fig:lognom}. $\mu$ and $\sigma$ are strongly correlated, with the peak match probability around $\mu\sim-2.7\equiv6.6\times10^{-14}$ erg s$^{-1}$ cm$^{-2}$ and $\sigma\sim1.7.$ The $p$-values are somewhat lower than in the power law analysis (this can be seen from 
\begin{figure}
\centerline{\includegraphics[width=3.2in]{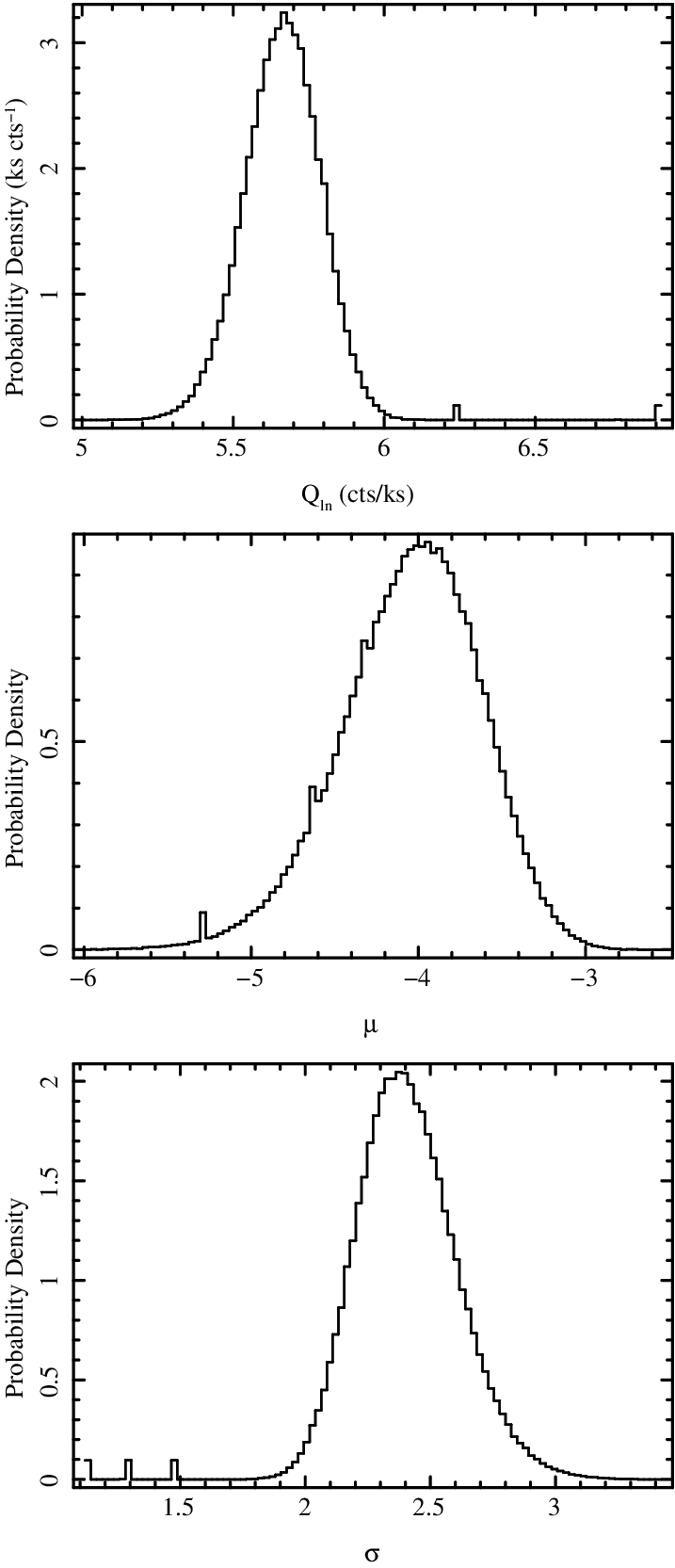}}
\caption{MCMC PDFs for the quiescent rate $Q_{ln}$ (top) and the log-normal location parameter $\mu$ (middle) and shape parameter $\sigma$ (bottom), marginalized over the other parameters. There are a few small islands of probability away from the peaks, but otherwise these distributions look similar to slightly skewed Gaussians. See text for details.\label{fig:lnpdf}}
\end{figure}
the contours themselves, which contain very little probability above the 68\% confidence line). While this result is suggestive, our likelihood analysis is better suited to determining the relative utility of the power law and log-normal models.

\subsubsection{Maximum Likelihood}
\label{sec:lnlike}
The MCMC analysis for the log-normal model is very similar to that of the power law model, although the likelihood function is somewhat more difficult to write down (Appendix \ref{sec:loglike}). Here, $a=3$ is sufficient to keep the acceptance fraction below 0.5.  Once again, we use 150 walkers per free parameter, this time initializing them to be uniformly distributed between $0.005\leq Q_{ln}$ cts$^{-1}$ s $\leq0.007,$ $-6\leq\mu\leq-2$ and $1\leq\sigma\leq4$. 

As the log-normal process typically contributes somewhat less flux (see Section \ref{sec:quiet}) than the power law process, the quiescent count rate $Q_{ln}$ is slightly higher than $Q_{pl}:~Q_{ln}=(5.7\pm0.1)\times10^{-3}$ cts s$^{-1}$. The log-normal component has a median flux $1.8_{-0.6}^{+0.8}\times10^{-14}$ erg s$^{-1}$ cm$^{-2}$ ($\mu=-4.0\pm0.4$) and a multiplicative standard deviation $\sigma=2.4\pm0.2.$ As noted above, most of the probability mass of the log-normal distribution lies below our detection limit.

The log-normal contours and PDFs are shown in black in Figures \ref{fig:lognom} and \ref{fig:lnpdf}. Although there was good qualitative and quantitative agreement between the MCMC and A-D results in the power law analysis, it does appear that these methods have different sensitivities: here there is much less overlap in the A-D contours and the MCMC contours: the synthetic data analysis prefers a higher median flux and a smaller multiplicative standard deviation, perhaps to reduce the number of simulated bins brighter than the highest-flux data point. If we fix $Q_{ln}=0.00526$ cts s$^{-1}$, the new MCMC contours lie between the A-D contours and the original MCMC contours, so it seems that the discrepancy is due in part to our choice to fix $Q_{ln}$ in the synthetic analysis. 

As in Section \ref{sec:pllike}, we draw random samples from the ($Q_{ln},~\mu,~\sigma$) ensemble, use them to generate simulated datasets, and plot the CDFs in the bottom panel of Figure \ref{fig:dist}. Again, the simulated data appear to be in fairly good agreement with the observed CDF over most of the observed count rates. The only noticeable difference with respect to the power law process is at the high flux end, where the data and the log-normal component diverge slightly. This is not entirely unexpected: the power law has an advantage over the log-normal in matching the turnover seen in the data at high flux (the former is bounded by definition, while the latter is not). A power law not bounded at the high flux end would behave comparably to the log-normal component here. Ultimately, however, this difference is significant enough to select the power law as a formally superior model: via the Akaike information criterion, due to the difference in log-likelihoods ($\sim6$) and the number of free parameters, the likelihood of the power law model relative to the log-normal model is $\gtrsim1300$. At face value, it seems that the log-normal model can effectively be ruled out for failing to perform as well as the power law at capturing the behavior of the very brightest flares from \sgra (but see Section \ref{sec:corr}).

\subsection{A Note on Correlations in the Data}
\label{sec:corr}
In this work, we are interested in the shape of the flux distribution (leaving the origin of that shape for future work), and we are able to reproduce the observed CDF using models and synthetic data with no explicit correlation. Although the well-studied flares are clearly indicative of correlated fluxes, our treatment here has precedent, as both \citetalias{Dodds-Eden11} and \citetalias{Witzel12} first analyzed the functional form of the NIR flux distribution without reference to correlations in the data. \citetalias{Witzel12} subsequently demonstrated that correlated synthetic light curves (generated using the structure function and power spectrum) provided a superior description of the flux distribution. For the moment, we note that our use of uncorrelated synthetic data to match the flux distribution does not necessarily imply that the actual fluxes are uncorrelated. In fact, our bounded power law model could be considered a proxy for a more comprehensive timing analysis (e.g., structure functions and damped random walks; \citetalias{Witzel12}; \citealt{Dexter14}), and the log-normal model should similarly not be considered complete (or definitively ruled out) until such correlations are accounted for. We will undertake this task in future work.



\subsection{Undetected Flares in Quiescence}
\label{sec:quiet}
\begin{figure}
\centerline{\includegraphics[width=3.2in]{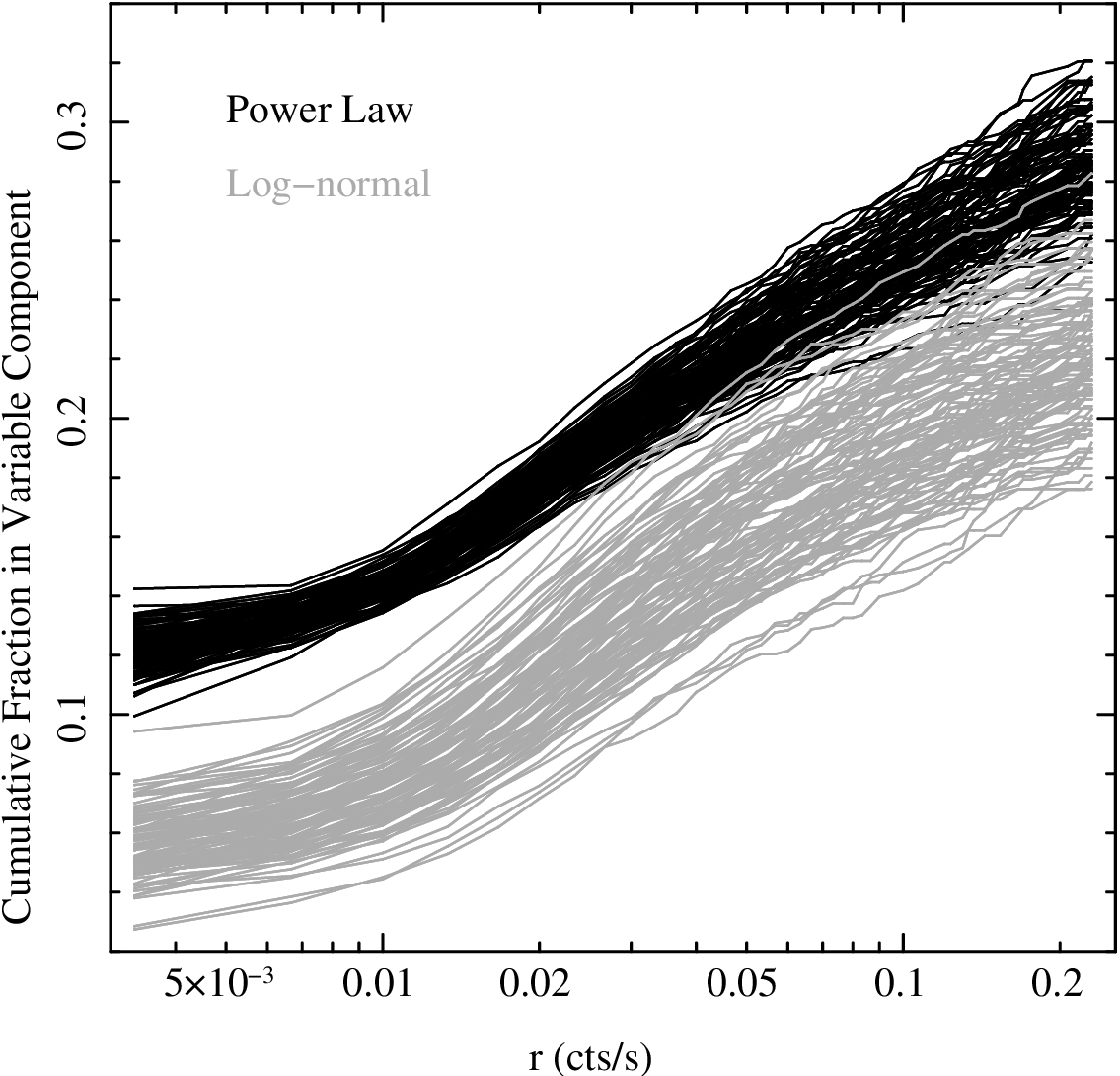}}
\caption{Mean cumulative fraction of the counts attributable to power law and log-normal processes as a function of count rate for 100 data sets drawn from the peak of the joint probability distributions. The power law typically contributes $\sim12\%$ in the lowest count rate bin, but contributes $\sim30\%$ of the total counts; the log-normal contributes a total of $\sim20-25\%$ of the counts. These contributions from the variable processes are as expected from our statistical analysis of the flares (i.e., a $\sim10\%$ contribution at low flux; \citealt{N13b}).\label{fig:frac}}
\end{figure}

We can also use our decomposition of the flux distribution into a steady component and a variable component to address the contribution of undetected flares to the quiescent emission from the inner $1.25\arcsec$ of the Galaxy. Here, we use the same random ($Q_{pl},~\xi$) and ($Q_{\ln},~\mu,~\sigma$) samples as in Figure \ref{fig:dist} to create synthetic data sets. For each data set, we calculate the fraction of the observed counts contributed by the variable process in each simulated time bin. In Figure \ref{fig:frac}, we show the cumulative fractional contributions of the power law and log-normal processes, i.e., the fraction of the counts at or below a given count rate that can be attributed to the variable component. 

Unsurprisingly, the low flux bins are dominated by the Poisson process. The power law process contributes a total of $\sim12\%$ of the counts in bins with one count and less than $20\%$ of the counts out to count rates of $\sim0.02$ cts s$^{-1}$ (corresponding to nearly $4\times$ the baseline quiescent count rate $Q_{pl}$). In general, the log-normal component contributes somewhat less than the power law, but the dependence on count rate is very similar. We may also note that the value of $Q_{pl}$ is only $\sim84\%$ of the average count rate outside the flares (\citealt{N13b}), so that the remaining flux must come from flares in our analysis. This accounting (i.e., a $\sim10-15\%$ contribution of undetected flares to the steady background emission) is consistent with the extrapolation of the observed flare distribution to low fluence and the power spectrum and the distribution of waiting times between photons in quiescence (\citealt{N13b}), as well as the quiescent spectrum (\citealt{Wang13}) and the surface brightness distribution of \sgra (\citealt{Shcherbakov10}).

\section{Radiation Mechanisms}
\label{sec:radiation}
In the preceding section, we demonstrate that the X-ray flux distribution of \sgra is consistent with a Poisson process, representing the steady thermal emission from the quiescent accretion flow, plus a power law process representing flare emission from within $\sim$tens of gravitational radii from the black hole (e.g., \citealt{Barriere14}). The flaring may also be consistent with a log-normal variability process, but the power law provides a superior description. As previous analyses have found evidence of a power law flux distribution in the NIR (\citetalias{Dodds-Eden11,Witzel12}), here we make a preliminary consideration of the extent to which a statistical comparison of the X-ray and NIR flux distributions can provide insight into the X-ray radiation mechanism. We proceed with the following formalism. The flux distribution (specifically, the differential flux distribution) is defined:
\begin{equation}
P(F_\lambda) \equiv \frac{dN}{dF_\lambda}, 
\end{equation}
and the cumulative flux distribution (i.e., the continuous extension of Equation \ref{eq:cdf}) is given by:\begin{equation}
N_{\geq}(F_\lambda) \equiv \int_{F_\lambda}^{\infty}P(f_\lambda)df_\lambda,
\label{eq:cum}
\end{equation}
where $F_\lambda$ is the flux at wavelength $\lambda$ and $dN/dF_\lambda$ is the number of times per unit flux that \sgra is observed with flux $F_\lambda$ at wavelength $\lambda$. Given an infrared flux distribution $P(F_{\rm IR})$, we may ask what any given radiation mechanism predicts for the X-ray flux distribution $P(F_{\rm X}).$ The IR is dominated by optically thin synchrotron emission (e.g., \citealt{Hornstein07}); for a one-zone model, the flux is:
\begin{equation}
\label{eq:synch}
F_\nu^{S}\propto n_eR^3B^{1+\alpha}\nu^{-\alpha},
\end{equation} 
where $n_e$ is the IR-emitting electron density, $R$ is the size of the emitting region, $B$ is the magnetic field, $\nu$ is the frequency, and $\alpha$ is the spectral index in the optically thin regime (e.g., \citealt{RL79}). For fluxes associated with detectable X-ray emission, we suppose the NIR flux distribution can be described as a power law with index $q,$ i.e., $P(F_{\rm IR})\propto F_{\rm IR}^{-q}.$ A parallel analysis could be performed to explore other functional forms for the flux distributions.

First, we derive a general result in the case of a power law relationship between the X-ray and IR, i.e., $F_{\rm X}\propto F_{\rm IR}^\beta,$ where $\beta\neq0$ is some constant. Under this condition, we have
\begin{eqnarray}
P(F_{\rm X}) &\equiv& \frac{dN}{dF_{\rm X}}\\
&\propto& F_{\rm IR}^{1-\beta}\frac{dN}{dF_{\rm IR}}\\
&\propto& F_{\rm IR}^{1-\beta-q}\\
&\propto& F_{\rm X}^{-(q+\beta-1)/\beta}.\label{eq:beta}
\end{eqnarray} In other words, if the infrared flux distribution is a power law with index $q$ and $F_{\rm X}\propto F_{\rm IR}^\beta$, the X-ray flux distribution will be a power law with index $\xi=(q+\beta-1)/\beta.$ It can also be seen that the index of the cumulative distribution in the X-ray is $1/\beta$ times that in the NIR.

In general, $\beta$ is not trivial to derive from first principles, even in a one-zone model. For instance, the synchrotron flux $F_{\rm IR}^{S}$ (see Equation \ref{eq:synch} above) depends on several physical parameters, and the response of the X-ray to any variations in $F_{\rm IR}$ will be sensitive to which of the synchrotron parameters vary, as well as their time dependence and the X-ray emission mechanism (see, e.g., \citealt{Dodds-Eden10,Yusef-Zadeh09,Eckart12}). This sort of detailed time-dependent modeling of individual flares is beyond the scope of this paper, and is furthermore not easily applied to a statistical analysis of the flux distribution.

However, it can still be enlightening to consider the multiwavelength flux distribution in this context. For a simple, illustrative example, let us consider a scenario in which the X-ray emission is the high-energy tail of the synchrotron radiation observed in the IR. Since the synchrotron cooling time at X-ray frequencies is much shorter than the duration of the flares, radiative losses are likely very important (\citealt{Kardashev62,Yuan03}). Indeed, modeling the spectral energy distribution of a single bright flare, \citet{Dodds-Eden09} found that synchrotron models required a cooling break (at an intermediate frequency $\nu_c\sim10^{15}$ Hz) in order to avoid violating their upper limits on the mid-IR flux. Such a break could also explain why the flare spectra may appear to be slightly steeper in the X-ray than in the NIR (c.f.\ \citealt{Hornstein07,Barriere14}). In this model, assuming that $\nu_c$ is in the optical or UV, the X-ray flux is given by
\begin{eqnarray}
F_{\rm X} &\approx& F_{\rm IR} \left(\frac{\nu_c}{\nu_{\rm IR}}\right)^{-\alpha}\left(\frac{\nu_{\rm X}}{\nu_c}\right)^{-(\alpha+1/2)}\\
&\propto& F_{\rm IR}~\nu_c^{1/2}.\label{eq:nuc}
\end{eqnarray}
If $\nu_c$ has a power-law dependence on the IR flux, then there is some $\beta$ for which $\nu_c\sim F_{\rm IR}^{2\beta -2}$, so that $F_{\rm X} \propto F_{\rm IR}^\beta$ and Equation \ref{eq:beta} applies.

Observationally, since $\xi\lesssim q,$ we use Equation \ref{eq:beta} to infer $\beta\gtrsim 1$. For the specific case of the cooling break in Equation \ref{eq:nuc}, $\beta\gtrsim 1$ implies that $\nu_c$ is either independent of or increases with $F_{\rm IR}.$ This is counterintuitive because $\nu_c$ and the synchrotron flux (Equation \ref{eq:synch}) decrease and increase, respectively, with both $B$ and $R$ (for $\nu_c,$ see, e.g., Equation 3 in \citealt{Dodds-Eden09}). However, $\beta\gtrsim1$ can be understood in the context of more sophisticated models of the time-dependent emission. For instance, \citet{Dodds-Eden10} modeled the light curve of a single flare with a temporary increase in $n_e$, anticorrelated with changes in the magnetic field (producing both an increase in flux \textit{and} $\beta>1$). Such correlations might be expected for magnetic reconnection events, and can be accounted for in self-consistent simulations of the plasma parameters during flares (e.g., \citealt{Dibi14a}; Dibi et al., in preparation). 

It is less straightforward to apply similar analytical considerations to the other proposed X-ray emission mechanisms (i.e., SSC and external Comptonization scenarios). Consider the SSC case. In the Thomson scattering limit for a homogeneous sphere of low optical depth (see \citealt{Bloom96,Marrone08} and references therein), the SSC flux from first-order scattering is proportional to the synchrotron flux times the Thomson optical depth, which is proportional to the product $n_e R$:
\begin{equation}
F_\nu^{\rm SSC}\propto n_e^2R^4B^{1+\alpha}\nu^{-\alpha}.\label{eq:ssc}
\end{equation} 
If the IR variability could be attributed to power law variations in either $n_e$ or $R$ in this scenario, one could write down $\beta$ and predict the X-ray flux distribution. However, we have already established that it is likely that flares involve possibly-correlated variations of several parameters. Furthermore, in the model of \citet{Dodds-Eden09}, the SSC flux peaked in or near the X-ray, so that the assumptions used to derive Equation \ref{eq:ssc} may not apply. A complete treatment, beyond the scope of this paper, would require integrating over the distributions of seed photons and electrons and allowing for variations in physical parameters. 

The situation for external Comptonization scenarios is similar. If the X-ray emission is produced by IR photons scattering off sub-mm-emitting electrons, we might expect the X-ray and IR to be related more or less linearly. If instead the sub-mm photons scatter off IR-emitting electrons, there need not be any correlation between the X-ray and infrared. However, the accuracy of these conclusions depends on the similarity of the plasma properties and evolution at sub-mm and IR wavelengths. 

\section{Conclusion}
\label{sec:discuss}
Perhaps it should come as no surprise that in the 15 years since \citet{Baganoff01,Baganoff03} discovered the incredibly faint counterpart of \sgra with \textit{Chandra}, it has proven difficult to pin down the underlying radiation physics of this elusive X-ray source. To date, there is still no consensus as to the X-ray emission mechanism or the underlying physical process(es) responsible for the variability. However, there appears to be some grounds for optimism, as the statistical behavior of the X-ray emission appears to mirror that of the near infrared, at least qualitatively. For instance, at both wavelengths, the source is continuously ``on" and its flux distribution is partially or fully consistent with a power law distribution  (e.g., \citealt{Dodds-Eden11,Witzel12,Meyer14_arxiv} and references therein for the NIR). 

Although no clear correlation has been found between the peak flux of flares observed simultaneously in NIR and X-rays (\citealt{Trap11}), the flare peaks are typically simultaneous within observational uncertainties (see \citealt{Yusef-Zadeh12} and references therein). Furthermore, flares at both wavelengths exhibit significant fast variations on short time scales (NIR: 47 s, \citealt{Dodds-Eden09}; X-ray: $\sim100$ s, \citealt{Nowak12,Barriere14}), indicating that excursions to high flux come from compact regions (within $\approx10$ Schwarzschild radii of the black hole in the case of the X-ray flares; \citealt{Barriere14}). The emission at different wavelengths need not be produced by the same electrons (e.g., \citealt{Hornstein07}), but it is plausible that these variable processes share a common physical origin.

Thus, the excellent data now available present a new and exciting opportunity to gain insight into the origin of the multiwavelength variability of \sgra. In this paper, we have explored the total X-ray flux distribution from the inner $1.25\arcsec$ of the Galaxy using statistical methods previously applied to \sgra's NIR emission. Rather than focusing on individual flares, we approach the radiation physics of \sgra from a different perspective, looking for connections between emission at different wavebands once we understand the statistical behavior of each waveband separately.

As is evident from Figure \ref{fig:dist}, the X-ray emission is dominated\footnote{Over time scales of 3 Ms. Since bright flares dominate the flare fluence (\citealt{N13b}), in principle one could simply observe until the integrated flare emission surpasses the quiescent source.} by the faint steady quiescent source and diffuse background. Our analysis of the flux distribution confirms the steadiness of the quiescent emission: roughly 85-90\% of it is consistent with a Poisson process, i.e., it exhibits no detectable intrinsic variability at all (see also \citealt{N13b}). To the extent that this thermal plasma emission from large scales (e.g., \citealt{Wang13}) is constant in time, all the X-ray variability from \sgra can be attributed to non-thermal flaring. 

The variable source, likely located in the inner accretion flow, is typically an order of magnitude fainter than the thermal plasma, but our statistical analysis indicates that it can be described by a single variability process over a dynamic range of three orders of magnitude in flux. At the faint end (below our flare detection limit), the variable source contributes the remaining $\sim10-15\%$ of the flux during quiescent/steady intervals; at the bright end, flares can exceed the steady background by factors of 100 or more (\citealt{Nowak12,Porquet03,Porquet08}). Both a power law process  (Equation \ref{eq:powlaw}) with index $\xi=1.92_{-0.02}^{+0.03}$ and a log-normal process (Equation \ref{eq:lognormal}) with location parameter $\mu=-4.0\pm0.4$ and shape parameter $\sigma=2.4\pm0.2$ are consistent with the observed excess above the quiescent emission, but the power law provides a significantly better description of the data. Notably, $\xi$ appears to be the same as the power law index of the distribution of flare luminosities ($1.9^{+0.3}_{-0.4}$; \citealt{N13b}). This result makes sense if the variable process is the sum of many discrete flares, but more work is required to know if this is a unique interpretation.

Our results place a strong observational constraint on theoretical models of the multiwavelength variability of \sgra. In addition to the flare SED and the variable X-ray/NIR ratio, any viable model must be able to reproduce the distribution of fluxes in both the NIR and the X-ray independently. Even the one-zone models considered in Section \ref{sec:radiation} can begin to inform our understanding of the relationship between the NIR and X-ray emission, but it is instructive to note why (in addition to the limited power of one-zone models) a complete answer may require more than simple analytical considerations. The primary reasons are physical:
\begin{itemize}
\item Even given an X-ray emission mechanism, the X-ray variability can only be predicted analytically from the NIR variability if the origin of the NIR variability is known. 
\item The NIR power law is probably not attributable to a single parameter with a power law distribution. Instead, the synchrotron flux likely varies due to non-power-law variations of multiple parameters (as in the time-dependent analysis of \citealt{Dodds-Eden10}).
\end{itemize}
Self-consistent calculations of the particle distribution during flares (\citealt{Dibi14a}) can overcome these difficulties, and we will analyze the multiwavelength flux distribution in the context of these calculations in future work (Dibi et al., in preparation). It may also be possible to understand the variability in light of other simulations of the accretion flow (e.g., \citealt{Moscibrodzka13}). 

In addition, several factors complicate multiwavelength statistical analyses of  \sgra:
\begin{itemize}
\item There is disagreement in the literature over the functional form of the NIR flux distribution, which leads to different estimates of the index $q$ of the power law portion (i.e., $q=2.7,$ \citetalias{Dodds-Eden11}; $q=4.2,$ \citetalias{Witzel12}). This is particularly relevant in the context of Equation \ref{eq:beta}.
\item Any deviations from a pure power law distribution break the scale-free nature of our calculations, so that in order to validate the comparison performed in Section \ref{sec:radiation}, we would need to know which NIR fluxes correspond to which X-ray fluxes. This may not be problematic, as deviations from a power law (\citealt{Dodds-Eden11}) have only been reported at NIR fluxes that are not typically associated with detectable X-ray flares ($\lesssim5$ mJy; see also \citealt{Marrone08}). 
\item The flux distribution may not be completely stationary on time scales of individual observations. \citetalias{Witzel12} argue that the two-state flux distribution of \citetalias{Dodds-Eden11} (i.e., lognormal at low flux, power law at high flux) is biased by a single bright flare. Such behavior might be expected if the power law is an aggregate effect of isolated flares in distinct regions of the accretion flow. Caution is therefore merited when comparing X-ray and NIR flux measurements taken over different time intervals and at different epochs (particularly because the X-ray flux distribution is consistent with several functional forms).
\item A related issue is that the characteristic time scales of the X-ray variability process are unknown. Although the near simultaneity of X-ray and NIR flares suggests that the two are fundamentally related, an empirical confirmation of similar variability time scales would lend significant credence to any multiwavelength statistical analysis of the radiation mechanisms at work in \sgra. We intend to perform this computationally-demanding measurement in the near future.
\end{itemize}

In short, there is still more work to be done to understand the shape of the flux distributions at different wavelengths and their relationships to each other. These points highlight the need for continued coordinated multiwavelength campaigns on \sgra (specifically, simultaneous observations). By tracking simultaneous variations in the X-ray and IR, we can make a direct measurement of the IR/X-ray flux scaling, free of any lingering concerns about observations taken at different epochs. Strictly simultaneous observations would also break the degeneracy related to the functional form of the NIR/X-ray flux distributions, since the same model or parameterization can be applied to both wavebands. Although complicated somewhat by frequency-dependent lags (e.g., \citealt{Marrone08,Falcke09} and references therein), the addition of radio/sub-mm variability statistical analyses (e.g., \citealt{Dexter14}) will paint a more comprehensive picture of variability in the Galactic center.

Although we must leave the question of the primary X-ray radiation mechanism of \sgra for future work, a statistical approach is still promising. Our clear and precise measurement of the X-ray flux distribution is a critical first step, and our analysis provides a new and strong observational constraint for theoretical models of \sgra. With the advances described above, the prospects for revealing the physics of variable X-ray emission in \sgra in the near future remain strong.

\acknowledgements We thank all the members of the \sgra \textit{Chandra} XVP collaboration, and we are immensely grateful to \textit{Chandra} Mission Planning for their support during our 2012 campaign. We thank Daniel Wang, Salom\'e Dibi, Feng Yuan, and Stefan Gillessen for very useful discussions concerning the analysis and interpretation of our data. We acknowledge the role of the Lorentz Center, Leiden, and the Netherlands Organization for Scientific Research Vidi Fellowship 639.042.711 (S.M.). J.N.\ gratefully acknowledges funding support from NASA through the Einstein Postdoctoral Fellowship, grant PF2-130097, awarded by the \textit{Chandra} X-ray Center, which is operated by the Smithsonian Astrophysical Observatory for NASA under contract NAS8-03060, and from NASA through the Smithsonian Astrophysical Observatory contract SV3-73016 to MIT for support of the \textit{Chandra} X-ray Center, which is operated by the Smithsonian Astrophysical Observatory for and on behalf of NASA under contract NAS8-03060. F.K.B.\ acknowledges support from the \textit{Chandra} grant G02-13110A under contract NAS8-03060. N.D.\ is supported by NASA through Hubble Postdoctoral Fellowship grant number HST-HF-51287.01-A. 

\appendix

\section{Poisson/Power Law Likelihood Function}
\label{sec:likelihood}
In order to use maximum likelihood methods, we naturally must be able to write down the likelihood function. In general, for the sum of two random variables $X$ and $Y$, the probability of observing a value of $n$ given parameters $\theta_X$ and $\theta_Y$ is a convolution:
\begin{equation}\mathcal{P}(n|\theta_X,\theta_Y)=\sum_{j=0}^{n}\mathcal{P}_X(j|\theta_X)\mathcal{P}_{Y}(n-j|\theta_Y),\label{eq:conv}
\end{equation} and for a dataset $D$ with $M$ measurements $n_i$, the likelihood is: 
\begin{equation}\mathcal{P}(D | \theta_X,\theta_Y)=\prod_{i=1}^{M}\mathcal{P}(n_i|\theta_X,\theta_Y).\label{eq:like}\end{equation} Below, we calculate the likelihood function using the models described in the main text.

As described in Section \ref{sec:pow}, we first model the light curve as the sum of a Poisson process and a power law process. The Poisson process has parameter $\Lambda\equiv(300~\textrm{s})~Q,$ where $Q$ is the quiescent count rate, and the well-known probability of observing $n$ counts is:
\begin{equation}\mathcal{P}_p(n|Q)=\frac{\Lambda^{n}e^{-\Lambda}}{n!}.\end{equation}
The power law probability is given by Equation \ref{eq:powlaw}, with $P(F)\propto F^{-\xi}$ on the interval $[F_1,~F_2].$ In order to use this power law in Equation \ref{eq:conv}, we must convert to counts, include the effects of pileup, and add counting noise. The latter two requirements make this a non-trivial likelihood function to write down.

First, we set out (and remind the reader of) some variables. As before, $n$ shall be the number of observed counts and $F$ shall be a power law flux. We shall use $c$ to denote the counts implied by applying the flux/counts scaling from the brightest flare (Section \ref{sec:flux}; see also \citealt{Nowak12}) to $F$, and $m$ shall be the result of applying pileup effects to $c.$ In this scheme, $F,$ $c,$ and $m$ are continuous variables, while $n$ is discrete. To add counting noise, we need to find the probability of observing $n$ counts given $m$ counts in the piled-up power law process and then integrate over all values of $m$: 
\begin{equation}\mathcal{P}_{pl}(n|\xi)=\int_{m_1}^{m_2}P(n|m)P(m|\xi)dm.\end{equation} Here $m_1$ and $m_2$ are the upper and lower limits of the integral calculated by scaling the bounds for the power law $F_1$ and $F_2$ to counts and then applying our pileup model. The complication is that $\mathcal{P}(m|\xi)$ is not a power law. Based on our analysis of pileup in these data (\citealt{Nowak12,N13b}), we have determined that $c(m)$ can be approximated to within 0.5\% over the observed range of count rates as:
\begin{equation}
c(m) = A_1 m^{A_2} e^{A_3 m},
\end{equation}
with $A_1=0.9761415, A_2=1.004234,$ and $A_3=0.002401043.$ Then 
\begin{eqnarray}
\mathcal{P}_{pl}(m|\xi) &=& \frac{dN}{dm}\\
&=& \frac{dN}{dc}\frac{dc}{dm}\\
&=&K_{pl}(A_1 m^{A_2} e^{A_3 m})^{1-\xi}(A_3 + A_2/m).\label{eq:dndm}
\end{eqnarray}

\begin{figure}[h]
\centerline{\includegraphics[width=4in,clip=true,trim=50 35 30 0]{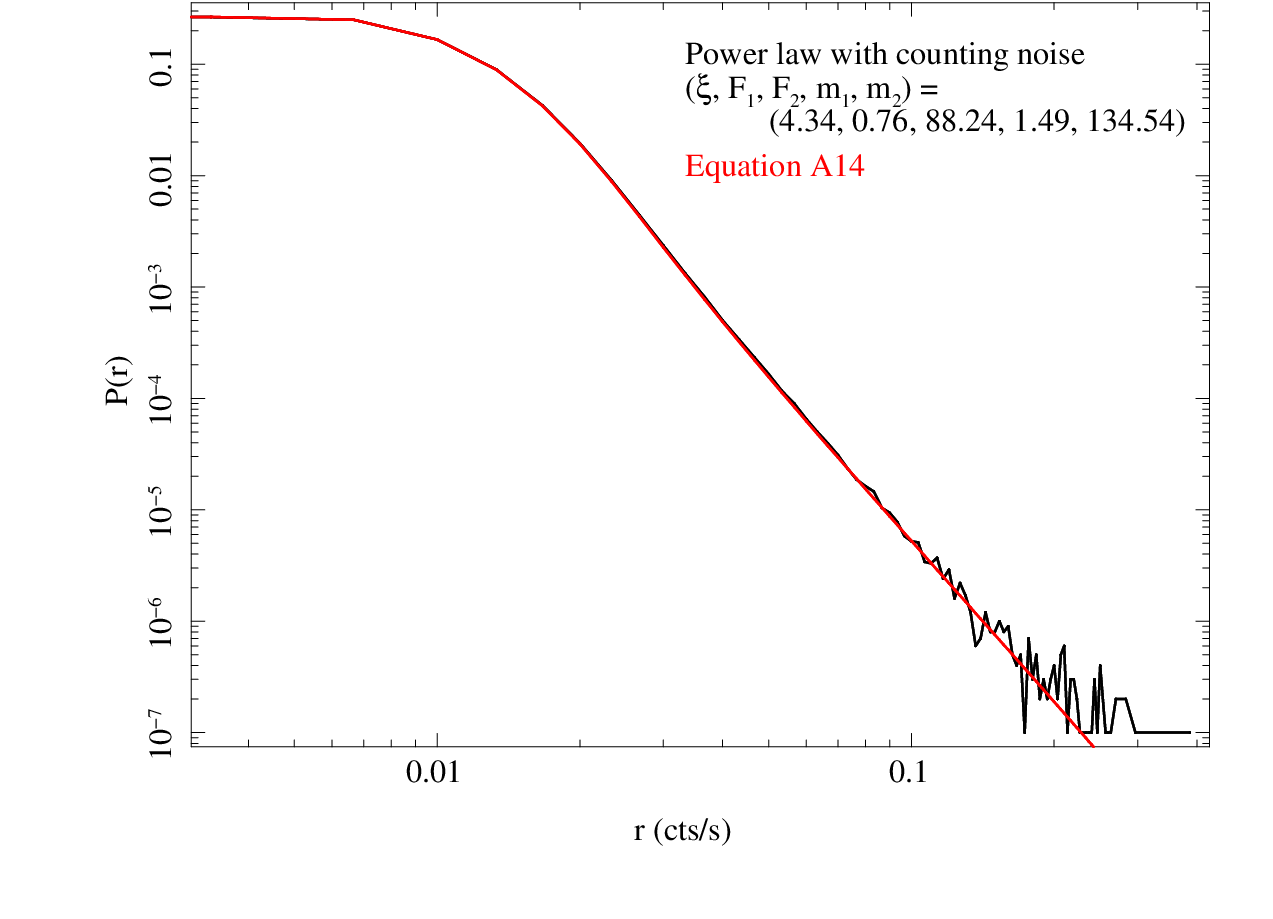}}
\caption{Probability of a given count rate for a randomly-selected power law with counting noise and pileup (black) and Equation \ref{eq:normp} (red). The random data are well reproduced by the formula.}\label{fig:prand}
\end{figure}
The normalization constant $K_{pl}$ is easily found to be:
\begin{equation}
K_{pl} = \frac{(\xi-1)}{(A_1 m_1^{A_2} e^{A_3 m_1})^{1-\xi}-(A_1 m_2^{A_2} e^{A_3 m_2})^{1-\xi}}.
\end{equation}

Finally, we can write the probability of observing $n$ counts in the power law as
\begin{eqnarray}
\mathcal{P}_{pl}(n|\xi) &\propto& K_{pl} \int_{m_1}^{m_2}\frac{m^n e^{-m}}{n!}(A_1 m^{A_2} e^{A_3 m})^{1-\xi}(A_3 + A_2/m) dm\\
&\equiv&\frac{K_{pl} A_1^{1-\xi}}{n!}\int_{m_1}^{m_2}m^{A-1} e^{-B m}(A_3 + A_2/m) dm\\
&=&\frac{K_{pl} A_1^{1-\xi}}{n!}\left[A_3 B^{-A}[\Gamma(A,m_1B)-\Gamma(A,m_2 B)]+A_2 B^{-A+1}[\Gamma(A-1,m_1B)-\Gamma(A-1,m_2 B)]\right] \\
&=&\frac{K_{pl} A_1^{1-\xi}B^{-A}}{n!}\left[A_3 [\Gamma(A,m_1B)-\Gamma(A,m_2 B)]+A_2 B[\Gamma(A-1,m_1B)-\Gamma(A-1,m_2 B)]\right].\label{eq:pnpl}
\end{eqnarray} For simplicity, we have made the substitutions $A \equiv n + A_2(1-\xi) + 1$ and $B=1-A_3(1-\xi).$ $\Gamma$ is the upper incomplete gamma function. It turns out that Equation \ref{eq:pnpl}  itself is not normalized, so we need divide by a sum over all plausible values of $n,$ \begin{equation}
\mathcal{P}_{pl}(n|\xi)\leftarrow \frac{\mathcal{P}_{pl}(n|\xi)}{\sum^{i=150}_{i=0}\mathcal{P}_{pl}(i|\xi)}.
\label{eq:normp}\end{equation}
To test this formula, we generated random numbers to represent $\xi,$ $F_1,$ and $F_2$ (4.34, 0.76, and 88.24, respectively), drew 10 million random numbers from the associated power law distribution, and incorporated pileup and counting noise. The distribution of the resulting counts is shown in black in Figure \ref{fig:prand}, and is very well matched by Equation \ref{eq:normp} (red). With confidence in the power law probability, we can easily use Equations \ref{eq:conv} and \ref{eq:like} to calculate the logarithm of the likelihood given any particular set of parameters.

\section{Poisson/Log-normal Likelihood Function}
\label{sec:loglike}

For the log-normal model, the situation is very similar, except for the obvious substitution in Equation \ref{eq:dndm}. Here, with the definition $\kappa=F/c,$ we have:
\begin{eqnarray}
\mathcal{P}_{ln}(m|\mu,\sigma)&=&\frac{dN}{dF}\frac{dF}{dc}\frac{dc}{dm}\\
&=&\frac{K_{ln}}{\sqrt{2\pi}\sigma F}\exp\left[-\frac{(\ln{F}-\mu)^2}{2\sigma^2}\right]\kappa A_1 m^{A_2}e^{A_3 m}\left(A_3+\frac{A_2}{m}\right)\\
&=&\frac{K_{ln}}{\sqrt{2\pi}\sigma}\exp\left[-\frac{(\ln{(\kappa A_1 m^{A_2}e^{A_3 m})}-\mu)^2}{2\sigma^2}\right]\left(A_3+\frac{A_2}{m}\right).
\end{eqnarray}

The normalization $K_{ln}$ can be found for a finite interval $[m_1,m_2]:$ 
\begin{equation}
2\left[{\rm erf}\left(\frac{\ln{(\kappa A_1 m_2^{A_2}e^{A_3 m_2})}-\mu}{\sqrt{2}\sigma}\right)-{\rm erf}\left(\frac{\ln{(\kappa A_1 m_1^{A_2}e^{A_3 m_1})}-\mu}{\sqrt{2}\sigma}\right)\right]^{-1}.
\end{equation}
In the limit $a\rightarrow0,~b\rightarrow\infty,~K_{ln}\approx1.$ Unfortunately, the probability of observing $n$ counts in the log-normal component, i.e.
\begin{equation}
\mathcal{P}_{ln}(n|\mu,\sigma) \propto \int_{m_1}^{m_2}\frac{m^n e^{-m}}{n!}\frac{K_{ln}}{\sqrt{2\pi}\sigma}\exp\left[-\frac{(\ln{(\kappa A_1 m^{A_2}e^{A_3 m})}-\mu)^2}{2\sigma^2}\right]\left(A_3+\frac{A_2}{m}\right) dm,
\end{equation} cannot be calculated analytically. Instead, we estimate $\mathcal{P}_{ln}(n|\mu,\sigma)$ empirically by generating 2.5 million log-normal random fluxes for each ($\mu,~\sigma$) pair, converting to piled-up counts, and adding counting noise. It is only necessary to estimate this quantity for $n$ between 0 and 69. The rest of the likelihood calculation proceeds as in Appendix \ref{sec:likelihood}.

\bibliographystyle{apj}
\bibliography{ms}

\begin{thebibliography}{}
\expandafter\ifx\csname natexlab\endcsname\relax\def\natexlab#1{#1}\fi

\bibitem[{{Anderson} \& {Darling}(1945)}]{AndersonDarling54}
{Anderson}, T.~W., \& {Darling}, D.~A. 1945, J.\ Am.\ Statist.\ Assoc.\, 49,
  765

\bibitem[{{Baganoff} {et~al.}(2001){Baganoff}, {Bautz}, {Brandt}, {Chartas},
  {Feigelson}, {Garmire}, {Maeda}, {Morris}, {Ricker}, {Townsley}, \&
  {Walter}}]{Baganoff01}
{Baganoff}, F.~K., {Bautz}, M.~W., {Brandt}, W.~N., {et~al.} 2001, \nat, 413,
  45

\bibitem[{{Baganoff} {et~al.}(2003){Baganoff}, {Maeda}, {Morris}, {Bautz},
  {Brandt}, {Cui}, {Doty}, {Feigelson}, {Garmire}, {Pravdo}, {Ricker}, \&
  {Townsley}}]{Baganoff03}
{Baganoff}, F.~K., {Maeda}, Y., {Morris}, M., {et~al.} 2003, \apj, 591, 891

\bibitem[{{Barri{\`e}re} {et~al.}(2014){Barri{\`e}re}, {Tomsick}, {Baganoff},
  {Boggs}, {Christensen}, {Craig}, {Dexter}, {Grefenstette}, {Hailey},
  {Harrison}, {Madsen}, {Mori}, {Stern}, {Zhang}, {Zhang}, \&
  {Zoglauer}}]{Barriere14}
{Barri{\`e}re}, N.~M., {Tomsick}, J.~A., {Baganoff}, F.~K., {et~al.} 2014,
  \apj, 786, 46

\bibitem[{{B{\'e}langer} {et~al.}(2005){B{\'e}langer}, {Goldwurm}, {Melia},
  {Ferrando}, {Grosso}, {Porquet}, {Warwick}, \& {Yusef-Zadeh}}]{Belanger05}
{B{\'e}langer}, G., {Goldwurm}, A., {Melia}, F., {et~al.} 2005, \apj, 635, 1095

\bibitem[{{Blandford} \& {Begelman}(1999)}]{Blandford99}
{Blandford}, R.~D., \& {Begelman}, M.~C. 1999, \mnras, 303, L1

\bibitem[{{Bloom} \& {Marscher}(1996)}]{Bloom96}
{Bloom}, S.~D., \& {Marscher}, A.~P. 1996, \apj, 461, 657

\bibitem[{{Clavel} {et~al.}(2013){Clavel}, {Terrier}, {Goldwurm}, {Morris},
  {Ponti}, {Soldi}, \& {Trap}}]{Clavel13}
{Clavel}, M., {Terrier}, R., {Goldwurm}, A., {et~al.} 2013, \aap, 558, A32

\bibitem[{{Davis}(2001)}]{Davis01}
{Davis}, J.~E. 2001, \apj, 562, 575

\bibitem[{{Degenaar} {et~al.}(2013){Degenaar}, {Miller}, {Kennea}, {Gehrels},
  {Reynolds}, \& {Wijnands}}]{Degenaar13}
{Degenaar}, N., {Miller}, J.~M., {Kennea}, J., {et~al.} 2013, \apj, 769, 155

\bibitem[{{Dexter} \& {Fragile}(2013)}]{Dexter13}
{Dexter}, J., \& {Fragile}, P.~C. 2013, \mnras, 432, 2252

\bibitem[{{Dexter} {et~al.}(2014){Dexter}, {Kelly}, {Bower}, {Marrone},
  {Stone}, \& {Plambeck}}]{Dexter14}
{Dexter}, J., {Kelly}, B., {Bower}, G.~C., {et~al.} 2014, \mnras, 442, 2797

\bibitem[{Dibi {et~al.}(2014)Dibi, Markoff, Belmont, Malzac, Barrire, \&
  Tomsick}]{Dibi14a}
Dibi, S., Markoff, S., Belmont, R., {et~al.} 2014, Monthly Notices of the Royal
  Astronomical Society, 441, 1005

\bibitem[{{Do} {et~al.}(2009){Do}, {Ghez}, {Morris}, {Yelda}, {Meyer}, {Lu},
  {Hornstein}, \& {Matthews}}]{Do09}
{Do}, T., {Ghez}, A.~M., {Morris}, M.~R., {et~al.} 2009, \apj, 691, 1021

\bibitem[{{Dodds-Eden} {et~al.}(2010){Dodds-Eden}, {Sharma}, {Quataert},
  {Genzel}, {Gillessen}, {Eisenhauer}, \& {Porquet}}]{Dodds-Eden10}
{Dodds-Eden}, K., {Sharma}, P., {Quataert}, E., {et~al.} 2010, \apj, 725, 450

\bibitem[{{Dodds-Eden} {et~al.}(2009){Dodds-Eden}, {Porquet}, {Trap},
  {Quataert}, {Haubois}, {Gillessen}, {Grosso}, {Pantin}, {Falcke}, {Rouan},
  {Genzel}, {Hasinger}, {Goldwurm}, {Yusef-Zadeh}, {Clenet}, {Trippe},
  {Lagage}, {Bartko}, {Eisenhauer}, {Ott}, {Paumard}, {Perrin}, {Yuan},
  {Fritz}, \& {Mascetti}}]{Dodds-Eden09}
{Dodds-Eden}, K., {Porquet}, D., {Trap}, G., {et~al.} 2009, \apj, 698, 676

\bibitem[{{Dodds-Eden} {et~al.}(2011){Dodds-Eden}, {Gillessen}, {Fritz},
  {Eisenhauer}, {Trippe}, {Genzel}, {Ott}, {Bartko}, {Pfuhl}, {Bower},
  {Goldwurm}, {Porquet}, {Trap}, \& {Yusef-Zadeh}}]{Dodds-Eden11}
{Dodds-Eden}, K., {Gillessen}, S., {Fritz}, T.~K., {et~al.} 2011, \apj, 728, 37

\bibitem[{{Eckart} {et~al.}(2004){Eckart}, {Baganoff}, {Morris}, {Bautz},
  {Brandt}, {Garmire}, {Genzel}, {Ott}, {Ricker}, {Straubmeier}, {Viehmann},
  {Sch{\"o}del}, {Bower}, \& {Goldston}}]{Eckart04}
{Eckart}, A., {Baganoff}, F.~K., {Morris}, M., {et~al.} 2004, \aap, 427, 1

\bibitem[{{Eckart} {et~al.}(2006){Eckart}, {Baganoff}, {Sch{\"o}del}, {Morris},
  {Genzel}, {Bower}, {Marrone}, {Moran}, {Viehmann}, {Bautz}, {Brandt},
  {Garmire}, {Ott}, {Trippe}, {Ricker}, {Straubmeier}, {Roberts},
  {Yusef-Zadeh}, {Zhao}, \& {Rao}}]{Eckart06}
{Eckart}, A., {Baganoff}, F.~K., {Sch{\"o}del}, R., {et~al.} 2006, \aap, 450,
  535

\bibitem[{{Eckart} {et~al.}(2009){Eckart}, {Baganoff}, {Morris}, {Kunneriath},
  {Zamaninasab}, {Witzel}, {Sch{\"o}del}, {Garc{\'{\i}}a-Mar{\'{\i}}n},
  {Meyer}, {Bower}, {Marrone}, {Bautz}, {Brandt}, {Garmire}, {Ricker},
  {Straubmeier}, {Roberts}, {Muzic}, {Mauerhan}, \& {Zensus}}]{Eckart09}
{Eckart}, A., {Baganoff}, F.~K., {Morris}, M.~R., {et~al.} 2009, \aap, 500, 935

\bibitem[{{Eckart} {et~al.}(2012){Eckart}, {Garc{\'{\i}}a-Mar{\'{\i}}n},
  {Vogel}, {Teuben}, {Morris}, {Baganoff}, {Dexter}, {Sch{\"o}del}, {Witzel},
  {Valencia-S.}, {Karas}, {Kunneriath}, {Straubmeier}, {Moser}, {Sabha},
  {Buchholz}, {Zamaninasab}, {Mu{\v z}i{\'c}}, {Moultaka}, \&
  {Zensus}}]{Eckart12}
{Eckart}, A., {Garc{\'{\i}}a-Mar{\'{\i}}n}, M., {Vogel}, S.~N., {et~al.} 2012,
  \aap, 537, A52

\bibitem[{{Falcke} {et~al.}(2004){Falcke}, {K{\"o}rding}, \&
  {Markoff}}]{Falcke04}
{Falcke}, H., {K{\"o}rding}, E., \& {Markoff}, S. 2004, \aap, 414, 895

\bibitem[{{Falcke} {et~al.}(2009){Falcke}, {Markoff}, \& {Bower}}]{Falcke09}
{Falcke}, H., {Markoff}, S., \& {Bower}, G.~C. 2009, \aap, 496, 77

\bibitem[{{Foreman-Mackey} {et~al.}(2013){Foreman-Mackey}, {Hogg}, {Lang}, \&
  {Goodman}}]{Foreman-Mackey13}
{Foreman-Mackey}, D., {Hogg}, D.~W., {Lang}, D., \& {Goodman}, J. 2013, \pasp,
  125, 306

\bibitem[{{Genzel} {et~al.}(2003){Genzel}, {Sch{\"o}del}, {Ott}, {Eckart},
  {Alexander}, {Lacombe}, {Rouan}, \& {Aschenbach}}]{Genzel03}
{Genzel}, R., {Sch{\"o}del}, R., {Ott}, T., {et~al.} 2003, \nat, 425, 934

\bibitem[{{Goldwurm} {et~al.}(2003){Goldwurm}, {Brion}, {Goldoni}, {Ferrando},
  {Daigne}, {Decourchelle}, {Warwick}, \& {Predehl}}]{Goldwurm03}
{Goldwurm}, A., {Brion}, E., {Goldoni}, P., {et~al.} 2003, \apj, 584, 751

\bibitem[{{Goodman} \& {Weare}(2010)}]{GoodmanWeare}
{Goodman}, J., \& {Weare}, J. 2010, Commun. Appl. Math. Comput. Sci, 5, 65

\bibitem[{{Hornstein} {et~al.}(2007){Hornstein}, {Matthews}, {Ghez}, {Lu},
  {Morris}, {Becklin}, {Rafelski}, \& {Baganoff}}]{Hornstein07}
{Hornstein}, S.~D., {Matthews}, K., {Ghez}, A.~M., {et~al.} 2007, \apj, 667,
  900

\bibitem[{{Houck}(2002)}]{Houck02}
{Houck}, J.~C. 2002, in High Resolution X-ray Spectroscopy with XMM-Newton and
  Chandra, ed. {G.~Branduardi-Raymont}

\bibitem[{{Houck} \& {Denicola}(2000)}]{HD00}
{Houck}, J.~C., \& {Denicola}, L.~A. 2000, in ASP Conf. Ser., Vol. 216,
  Astronomical Data Analysis Software and Systems IX, ed. {N.~Manset,
  C.~Veillet, \& D.~Crabtree} (San Francisco, CA: ASP), 591

\bibitem[{{Kardashev} {et~al.}(1962){Kardashev}, {Kuz'min}, \&
  {Syrovatskii}}]{Kardashev62}
{Kardashev}, N.~S., {Kuz'min}, A.~D., \& {Syrovatskii}, S.~I. 1962, \sovast, 6,
  167

\bibitem[{{Kosti{\'c}} {et~al.}(2009){Kosti{\'c}}, {{\v C}ade{\v z}},
  {Calvani}, \& {Gomboc}}]{Kostic09}
{Kosti{\'c}}, U., {{\v C}ade{\v z}}, A., {Calvani}, M., \& {Gomboc}, A. 2009,
  \aap, 496, 307

\bibitem[{{Liu} \& {Melia}(2002)}]{Liu02}
{Liu}, S., \& {Melia}, F. 2002, \apjl, 566, L77

\bibitem[{{Liu} {et~al.}(2004){Liu}, {Petrosian}, \& {Melia}}]{Liu04}
{Liu}, S., {Petrosian}, V., \& {Melia}, F. 2004, \apjl, 611, L101

\bibitem[{{Markoff} {et~al.}(2001){Markoff}, {Falcke}, {Yuan}, \&
  {Biermann}}]{Markoff01}
{Markoff}, S., {Falcke}, H., {Yuan}, F., \& {Biermann}, P.~L. 2001, \aap, 379,
  L13

\bibitem[{{Markoff} {et~al.}(2005){Markoff}, {Nowak}, \& {Wilms}}]{Markoff05}
{Markoff}, S., {Nowak}, M.~A., \& {Wilms}, J. 2005, \apj, 635, 1203

\bibitem[{{Marrone} {et~al.}(2008){Marrone}, {Baganoff}, {Morris}, {Moran},
  {Ghez}, {Hornstein}, {Dowell}, {Mu{\~n}oz}, {Bautz}, {Ricker}, {Brandt},
  {Garmire}, {Lu}, {Matthews}, {Zhao}, {Rao}, \& {Bower}}]{Marrone08}
{Marrone}, D.~P., {Baganoff}, F.~K., {Morris}, M.~R., {et~al.} 2008, \apj, 682,
  373

\bibitem[{{McHardy} {et~al.}(2006){McHardy}, {Koerding}, {Knigge}, {Uttley}, \&
  {Fender}}]{McHardy06}
{McHardy}, I.~M., {Koerding}, E., {Knigge}, C., {Uttley}, P., \& {Fender},
  R.~P. 2006, \nat, 444, 730

\bibitem[{{Merloni} {et~al.}(2003){Merloni}, {Heinz}, \& {di
  Matteo}}]{Merloni03}
{Merloni}, A., {Heinz}, S., \& {di Matteo}, T. 2003, \mnras, 345, 1057

\bibitem[{{Meyer} {et~al.}(2008){Meyer}, {Do}, {Ghez}, {Morris}, {Witzel},
  {Eckart}, {B{\'e}langer}, \& {Sch{\"o}del}}]{Meyer08}
{Meyer}, L., {Do}, T., {Ghez}, A., {et~al.} 2008, \apjl, 688, L17

\bibitem[{{Meyer} {et~al.}(2009){Meyer}, {Do}, {Ghez}, {Morris}, {Yelda},
  {Sch{\"o}del}, \& {Eckart}}]{Meyer09}
---. 2009, \apjl, 694, L87

\bibitem[{{Meyer} {et~al.}(2014){Meyer}, {Witzel}, {Longstaff}, \&
  {Ghez}}]{Meyer14_arxiv}
{Meyer}, L., {Witzel}, G., {Longstaff}, F.~A., \& {Ghez}, A.~M. 2014, ArXiv
  e-prints, arXiv:1403.5289

\bibitem[{{Mo{\'s}cibrodzka} \& {Falcke}(2013)}]{Moscibrodzka13}
{Mo{\'s}cibrodzka}, M., \& {Falcke}, H. 2013, \aap, 559, L3

\bibitem[{{Narayan} {et~al.}(1995){Narayan}, {Yi}, \& {Mahadevan}}]{Narayan95a}
{Narayan}, R., {Yi}, I., \& {Mahadevan}, R. 1995, \nat, 374, 623

\bibitem[{{Neilsen} {et~al.}(2013){Neilsen}, {Nowak}, {Gammie}, {Dexter},
  {Markoff}, {Haggard}, {Nayakshin}, {Wang}, {Grosso}, {Porquet}, {Tomsick},
  {Degenaar}, {Fragile}, {Houck}, {Wijnands}, {Miller}, \& {Baganoff}}]{N13b}
{Neilsen}, J., {Nowak}, M.~A., {Gammie}, C., {et~al.} 2013, \apj, 774, 42

\bibitem[{{Nowak} {et~al.}(2012){Nowak}, {Neilsen}, {Markoff}, {Baganoff},
  {Porquet}, {Grosso}, {Levin}, {Houck}, {Eckart}, {Falcke}, {Ji}, {Miller}, \&
  {Wang}}]{Nowak12}
{Nowak}, M.~A., {Neilsen}, J., {Markoff}, S.~B., {et~al.} 2012, \apj, 759, 95

\bibitem[{{Pettitt}(1976)}]{Pettitt76}
{Pettitt}, A.~N. 1976, Biometrika, 63, 161

\bibitem[{{Plotkin} {et~al.}(2012){Plotkin}, {Markoff}, {Kelly}, {K{\"o}rding},
  \& {Anderson}}]{Plotkin12}
{Plotkin}, R.~M., {Markoff}, S., {Kelly}, B.~C., {K{\"o}rding}, E., \&
  {Anderson}, S.~F. 2012, \mnras, 419, 267

\bibitem[{{Ponti} {et~al.}(2010){Ponti}, {Terrier}, {Goldwurm}, {Belanger}, \&
  {Trap}}]{Ponti10}
{Ponti}, G., {Terrier}, R., {Goldwurm}, A., {Belanger}, G., \& {Trap}, G. 2010,
  \apj, 714, 732

\bibitem[{{Porquet} {et~al.}(2003){Porquet}, {Predehl}, {Aschenbach}, {Grosso},
  {Goldwurm}, {Goldoni}, {Warwick}, \& {Decourchelle}}]{Porquet03}
{Porquet}, D., {Predehl}, P., {Aschenbach}, B., {et~al.} 2003, \aap, 407, L17

\bibitem[{{Porquet} {et~al.}(2008){Porquet}, {Grosso}, {Predehl}, {Hasinger},
  {Yusef-Zadeh}, {Aschenbach}, {Trap}, {Melia}, {Warwick}, {Goldwurm},
  {B{\'e}langer}, {Tanaka}, {Genzel}, {Dodds-Eden}, {Sakano}, \&
  {Ferrando}}]{Porquet08}
{Porquet}, D., {Grosso}, N., {Predehl}, P., {et~al.} 2008, \aap, 488, 549

\bibitem[{{Quataert}(2002)}]{Quataert02}
{Quataert}, E. 2002, \apj, 575, 855

\bibitem[{{Quataert} \& {Gruzinov}(2000)}]{Quataert00a}
{Quataert}, E., \& {Gruzinov}, A. 2000, \apj, 539, 809

\bibitem[{{Rybicki} \& {Lightman}(1979)}]{RL79}
{Rybicki}, G.~B., \& {Lightman}, A.~P. 1979, {Radiative processes in
  astrophysics}

\bibitem[{{Scholz} \& {Stephens}(1987)}]{Scholz87}
{Scholz}, F.~W., \& {Stephens}, M.~A. 1987, J.\ Am.\ Statist.\ Assoc.\, 82, 918

\bibitem[{{Shcherbakov} \& {Baganoff}(2010)}]{Shcherbakov10}
{Shcherbakov}, R.~V., \& {Baganoff}, F.~K. 2010, \apj, 716, 504

\bibitem[{{Trap} {et~al.}(2011){Trap}, {Goldwurm}, {Dodds-Eden}, {Weiss},
  {Terrier}, {Ponti}, {Gillessen}, {Genzel}, {Ferrando}, {B{\'e}langer},
  {Cl{\'e}net}, {Rouan}, {Predehl}, {Capelli}, {Melia}, \&
  {Yusef-Zadeh}}]{Trap11}
{Trap}, G., {Goldwurm}, A., {Dodds-Eden}, K., {et~al.} 2011, \aap, 528, A140

\bibitem[{{Uttley} \& {McHardy}(2005)}]{Uttley05}
{Uttley}, P., \& {McHardy}, I.~M. 2005, \mnras, 363, 586

\bibitem[{{{\v C}ade{\v z}} {et~al.}(2008){{\v C}ade{\v z}}, {Calvani}, \&
  {Kosti{\'c}}}]{Cadez08}
{{\v C}ade{\v z}}, A., {Calvani}, M., \& {Kosti{\'c}}, U. 2008, \aap, 487, 527

\bibitem[{{Wang} {et~al.}(2013){Wang}, {Nowak}, {Markoff}, {Baganoff},
  {Nayakshin}, {Yuan}, {Cuadra}, {Davis}, {Dexter}, {Fabian}, {Grosso},
  {Haggard}, {Houck}, {Ji}, {Li}, {Neilsen}, {Porquet}, {Ripple}, \&
  {Shcherbakov}}]{Wang13}
{Wang}, Q.~D., {Nowak}, M.~A., {Markoff}, S.~B., {et~al.} 2013, Science, 341,
  981

\bibitem[{{Witzel} {et~al.}(2012){Witzel}, {Eckart}, {Bremer}, {Zamaninasab},
  {Shahzamanian}, {Valencia-S.}, {Sch{\"o}del}, {Karas}, {Lenzen}, {Marchili},
  {Sabha}, {Garcia-Marin}, {Buchholz}, {Kunneriath}, \&
  {Straubmeier}}]{Witzel12}
{Witzel}, G., {Eckart}, A., {Bremer}, M., {et~al.} 2012, \apjs, 203, 18

\bibitem[{{Xu} {et~al.}(2006){Xu}, {Narayan}, {Quataert}, {Yuan}, \&
  {Baganoff}}]{Xu06}
{Xu}, Y.-D., {Narayan}, R., {Quataert}, E., {Yuan}, F., \& {Baganoff}, F.~K.
  2006, \apj, 640, 319

\bibitem[{{Yuan} {et~al.}(2012){Yuan}, {Bu}, \& {Wu}}]{Yuan12}
{Yuan}, F., {Bu}, D., \& {Wu}, M. 2012, \apj, 761, 130

\bibitem[{{Yuan} {et~al.}(2009){Yuan}, {Lin}, {Wu}, \& {Ho}}]{Yuan09b}
{Yuan}, F., {Lin}, J., {Wu}, K., \& {Ho}, L.~C. 2009, \mnras, 395, 2183

\bibitem[{{Yuan} \& {Narayan}(2014)}]{Yuan14}
{Yuan}, F., \& {Narayan}, R. 2014, ArXiv e-prints, arXiv:1401.0586

\bibitem[{{Yuan} {et~al.}(2003){Yuan}, {Quataert}, \& {Narayan}}]{Yuan03}
{Yuan}, F., {Quataert}, E., \& {Narayan}, R. 2003, \apj, 598, 301

\bibitem[{{Yuan} {et~al.}(2004){Yuan}, {Quataert}, \& {Narayan}}]{Yuan04}
---. 2004, \apj, 606, 894

\bibitem[{{Yusef-Zadeh} {et~al.}(2009){Yusef-Zadeh}, {Bushouse}, {Wardle},
  {Heinke}, {Roberts}, {Dowell}, {Brunthaler}, {Reid}, {Martin}, {Marrone},
  {Porquet}, {Grosso}, {Dodds-Eden}, {Bower}, {Wiesemeyer}, {Miyazaki}, {Pal},
  {Gillessen}, {Goldwurm}, {Trap}, \& {Maness}}]{Yusef-Zadeh09}
{Yusef-Zadeh}, F., {Bushouse}, H., {Wardle}, M., {et~al.} 2009, \apj, 706, 348

\bibitem[{{Yusef-Zadeh} {et~al.}(2012){Yusef-Zadeh}, {Wardle}, {Dodds-Eden},
  {Heinke}, {Gillessen}, {Genzel}, {Bushouse}, {Grosso}, \&
  {Porquet}}]{Yusef-Zadeh12}
{Yusef-Zadeh}, F., {Wardle}, M., {Dodds-Eden}, K., {et~al.} 2012, \aj, 144, 1

\bibitem[{{Zubovas} {et~al.}(2012){Zubovas}, {Nayakshin}, \&
  {Markoff}}]{Zubovas12}
{Zubovas}, K., {Nayakshin}, S., \& {Markoff}, S. 2012, \mnras, 421, 1315

\end{thebibliography}

\label{lastpage}

\end{document}